\documentclass[conference]{IEEEtran}
\IEEEoverridecommandlockouts
\usepackage{fancyhdr}
\IEEEoverridecommandlockouts
\usepackage{fancyhdr}
\usepackage{cite}
\usepackage{amsmath,amssymb,amsfonts}
\usepackage{algorithmic}
\usepackage{graphicx}
\usepackage{textcomp}
\usepackage{xcolor}
\usepackage{array}    
\usepackage{fancyhdr}
\def\BibTeX{{\rm B\kern-.05em{\sc i\kern-.025em b}\kern-.08em
    T\kern-.1667em\lower.7ex\hbox{E}\kern-.125emX}}

\usepackage{titlesec}

\usepackage{amsmath}

\renewcommand{\thesection}{\arabic{section}}
\renewcommand{\thesubsection}{\arabic{section}.\arabic{subsection}}
\renewcommand{\thesubsubsection}{\arabic{section}.\arabic{subsection}.\arabic{subsubsection}}

\titleformat{\section}[block]{\bfseries\normalsize}{\thesection\quad}{0pt}{}  
\titleformat{\subsection}[block]{\itshape\normalsize}{\thesubsection\quad}{0pt}{}  
\titleformat{\subsubsection}[block]{\itshape\normalsize}{\thesubsubsection\quad}{0pt}{} 

\usepackage{float}
\usepackage[english]{babel}
\usepackage{amsthm}
\newtheorem{definition}{Definition}

\usepackage{multirow} 
\usepackage{booktabs}

\newtheorem*{remark}{Remark}

\begin{document}
\rmfamily

\title{{\fontsize{16}{19}\selectfont \textbf{Fairness-aware Dynamic Hosting Capacity and \\ the Impacts of Strategic Solar PV Curtailment}}} 


\author{

    \IEEEauthorblockN{{\fontsize{12}{14}\selectfont Beyzanur Aydin}}
    \IEEEauthorblockA{\textit{\fontsize{10}{12}\selectfont Department of EBE} \\
    \textit{\fontsize{10}{12}\selectfont University of Vermont}\\
    \fontsize{10}{12}\selectfont Burlington, USA \\
    \fontsize{10}{12}\selectfont beyzanur.aydin@uvm.edu}
    \and
   
    \IEEEauthorblockN{{\fontsize{12}{14}\selectfont Rebecca Holt}}
    \IEEEauthorblockA{\textit{\fontsize{10}{12}\selectfont Department of EBE} \\
    \textit{\fontsize{10}{12}\selectfont University of Vermont}\\
    \fontsize{10}{12}\selectfont Burlington, USA \\
    \fontsize{10}{12}\selectfont rebeccaholt581@gmail.com}
    \and
   
    \IEEEauthorblockN{{\fontsize{12}{14}\selectfont Mads R. Almassalkhi}}
    \IEEEauthorblockA{\textit{\fontsize{10}{12}\selectfont Department of EBE} \\
    \textit{\fontsize{10}{12}\selectfont University of Vermont}\\
    \fontsize{10}{12}\selectfont Burlington, USA \\
    \fontsize{10}{12}\selectfont malmassa@uvm.edu}
}

\maketitle
\thispagestyle{fancy}

\begin{abstract}
Rapid deployment of distributed energy resources (DERs), such as solar photovoltaics (PV), poses a risk to the distribution grid under high penetration. Therefore, studying hosting capacity (HC) limits considering grid physics and demand variability is crucial. This paper introduces an improved framework for determining the HC of radial distribution networks by enhancing an existing convex inner approximation (CIA) approach. The proposed method achieves a more accurate and larger inner approximation, resulting in better HC limits. We also consider time-varying demand and the design of objective functions to ensure equitable access to grid resources. A case study with solar PV integration is conducted using a modified IEEE-37 radial network to examine the impact of increased PV capacity, demonstrating that with no more than 5\% annual solar PV energy curtailed, it is possible to increase solar PV hosting capacity by at least 50\% with no negative grid impacts and a net positive economic impact when accounted for the cost of carbon.
Results show that fair allocation methods can lead to higher net profits and reduced PV curtailment and CO$_2$.
\end{abstract}


\IEEEpeerreviewmaketitle

\section{Introduction}
To support grid decarbonization by 2035, the US solar industry is expected to install at least 40~GW annually from 2025 onwards, with a projected average growth of 4\% per year~\cite{seia2024solar}. This growth is largely driven by the significant decline in the levelized cost of electricity (LCOE) for solar photovoltaics (PV), along with technological advancements and supportive policies that are increasing the adoption of solar PV~\cite{AEO2023}. However, without proper management and planning, these newly added assets could introduce operational challenges to the distribution grid, such as voltage regulation issues or power quality degradations caused by reverse power flows~\cite{HAQUE20161195}. 
Due to the uncertain nature of solar PV generation, especially at high penetration levels, active power curtailment can be used to regulate voltages~\cite{Bletterie2017,badmus2024anocaacnetworkawareoptimal}. The prevailing approach in renewable energy integration is to avoid curtailing surplus energy, as curtailing represents a loss of clean energy generation. To reduce curtailment, utilities have adopted market-based strategies, such as economic dispatch, negative pricing, and the energy imbalance market~\cite{bird2014wind}. In the US, these strategies are complemented by other solutions, including the deployment of energy storage and the expansion of transmission capacity~\cite{AEO2023}. However, instead of preventing PV curtailment altogether, recent methods seek to manage and optimize PV curtailment as a source of flexibility~\cite{Curt2020}.
But PV curtailment should only be used when the grid is otherwise at or beyond its hosting capacity (HC) limit. Thus, to ensure reliability in a high PV future, it is valuable to determine the limits of the hosting capacity of the grid. 

To address reliability concerns, utilities today often limit the deployment of solar PV by assigning conservative HC limits. These HC limits are typically static, and in reality, there is often potential to accommodate more PV capacity without affecting grid reliability. Once the static limit is reached, due to the first-come, first-served approach, the last customer attempting to install additional PV is either required to bear the costs of necessary grid upgrades or is denied installation. This situation limits PV deployments, raises concerns with decarbonization, and constrains consumer choice. Therefore, understanding the grid's hosting capacity limits is important to maximize the renewable energy potential of the grid.

To calculate the HC limits, optimization-based methods are employed in the literature where the objective function is to maximize the active power injections with respect to grid constraints. In this perspective, optimal power flow (OPF)-based techniques that consider the underlying grid physics stand out to provide reliable operation when determining the limits of the network~\cite{9640143, OPF_based_OE}. However, due to the non-linear nature of AC OPF equations, the problem becomes non-convex. To address this non-convexity, various methods such as convex relaxations and linearization techniques have been proposed in the literature. Convex relaxation techniques can sometimes fail to provide exact solutions when dealing with non-convex problems~\cite{CVX_relax_fail}. Similarly, linear approximations can lead to inaccuracies, such as over- or under-voltages, when applied to power system models~\cite{NAWAF2019}. Additionally, in practice, the priority is often placed on guaranteeing network admissibility rather than solely focusing on achieving the globally optimal solution. 

In~\cite{NAWAF2021}, authors introduce a novel convex inner approximation (CIA) method for the AC OPF problem, aimed at determining the admissible range of distributed energy resources (DERs) nodal injections in radial, balanced distribution feeders. This formulation leverages the \textit{DistFlow} equations~\cite{baran1989} to represent the physics of radial networks and employs convex envelopes to address the non-convexity introduced by the branch current equations. Within the set of admissible injections defined by a hyperrectangular set using CIA, safe and reliable operation of DERs is ensured while allowing DERs to operate independently. This paper builds upon and extends the work presented in~\cite{NAWAF2021} by improving the system model with tighter convex envelopes, thereby improving the HC analysis.

Besides, hosting capacity calculation is essentially a resource allocation problem, which makes incorporating notions of justice into energy system design inevitable. This leads to the necessity of ensuring equitable access to energy resources to be fair among all customers. In reality, it is seen that DER deployment is not distributed equitably among customers, prompting federal and state programs to support equitable access~\cite{Brockway2021, OShaughnessy2021}.

In this paper, we evaluate the equity of the determined nodal capacities from two perspectives: spatial fairness and temporal fairness. Spatial fairness refers to the capacity differences arising from the location of DERs, reflecting disparities caused by the physics of the grid. Temporal fairness considers the consistency of the HC limits over time; for instance, avoiding scenarios where the HC is very high at time $t$ and drops to zero at time $t + \Delta t$.
To address fairness, in many resource allocation or HC studies, logarithms are being used to ensure fairness among different nodes~\cite{NAWAF2021, Liu2024, Petrou2020, ogryczak2003}. However, the fairness of these methods has not been quantified or analyzed, or some have primarily been evaluated within the context of a single time step. In this paper, we present a case study highlighting the inequity that arises when maximizing hosting capacity without fairness considerations.We demonstrate how the objective function design effects this issue. To explicitly constraint a minimum fairness in decision-making, we incorporate the convex approach from~\cite{sundar2024parametricsecondorderconerepresentable}  within the optimization problem itself. Furthermore, we study the fairness of the hosting capacity allocations across time and space.
Additionally, the framework is enhanced to account for solar PV curtailment, assess associated carbon footprint impacts, and enable equitable grid access in solar PV integration scenarios.
The primary contributions of this research include:
 
\begin{enumerate}
   
    \item
    We introduce tighter convex envelopes for branch currents in the convex inner approximation method proposed in~\cite{NAWAF2021}, which improves the HC.
    
    \item Develop a methodology to constrain and analyze fairness in hosting capacity allocations, considering both temporal and spatial dimensions.
    
    \item Through a simulation-based study, we demonstrate the potential benefits of a small amount of solar PV curtailment in achieving significantly higher hosting capacity by coordinating solar PV curtailment. 
    \item Then we quantify the resulting carbon benefits of coordinating solar PV curtailment equitably, highlighting increased societal carbon savings, and demonstrate the importance of considering local energy profiles when evaluating the environmental impacts of DER integration.
   
\end{enumerate}

The remainder of the paper is structured as follows: Section~\ref{sec:section_CIA} defines the system model and presents a motivating example that emphasizes the significance of determining HC limits, followed by an explanation of the CIA methodology, including enhancements to the upper bound on branch currents. Section~\ref{sec:Fairness_of_CIA} analyzes one of the challenges associated with determining the HC of the grid and quantifies the fairness of allocation of resources. Section~\ref{sec:PV_Curtailment} presents a case study of CIA to further improve the limits of solar hosting capacity (SHC) and simulates potential carbon benefits associated with the increased solar PV. Section~\ref{sec:conclusion} concludes this paper and points out the future directions.
 

\section{Modeling Balanced Feeders} 
\label{sec:section_CIA}

We use \textit{DistFlow} equations to model the balanced single-phase radial distribution network~\cite{baran1989}.
\subsection{Modeling the Radial Network}
The radial distribution network is considered an undirected graph $\mathcal{G} := \{\mathcal{N} \cup \{0\}, \mathcal{L}\}$ that consists of $N + 1$ nodes and $L$ branches where $\mathcal{N} := \{1, \dots, N\}$ and the set of branches between nodes $i$ and $j$ is defined as $ \mathcal{L} := \{1, \dots, L\} \subseteq \mathcal{N} \times \mathcal{N}$, such that $(i, j) \in \mathcal{L}$. Node 0 is the substation bus with a fixed voltage $V_0$. Let $E \in \mathbb{R}^{(N + 1) \times N}$ be the incidence matrix of graph $\mathcal{G}$ relating the branches in $\mathcal{L}$ in between $\mathcal{N} \cup \{0\}$. $E$ has entry of 1 if there is a connection between $(i,j)$-th buses and, otherwise 0. The \textit{DistFlow} formulation uses squared voltages and branch currents; thus, denote $v_i := |V_i|^2$ as the voltage phasor at node $i \in \mathcal{N}$, and $l_{ij} := |I_{ij}|^2$ is the branch current phasor in line $(i,j) \in \mathcal{L}$. The active and reactive power flowing from bus $i$ to $j$ is $P_{ij}$ and $Q_{ij}$, respectively. The impedance ($z_{ij}$), resistance ($r_{ij}$), and reactance ($x_{ij}$) of each line $(i, j)$ is given as $z_{ij} := r_{ij} + \textbf{j}x_{ij}$ \footnote{Inductive if $x_{ij} > 0$ and capacitive if $x_{ij} < 0$.}.
\begin{figure}[!htbp]
    \centering
    \includegraphics[width=0.6\linewidth]{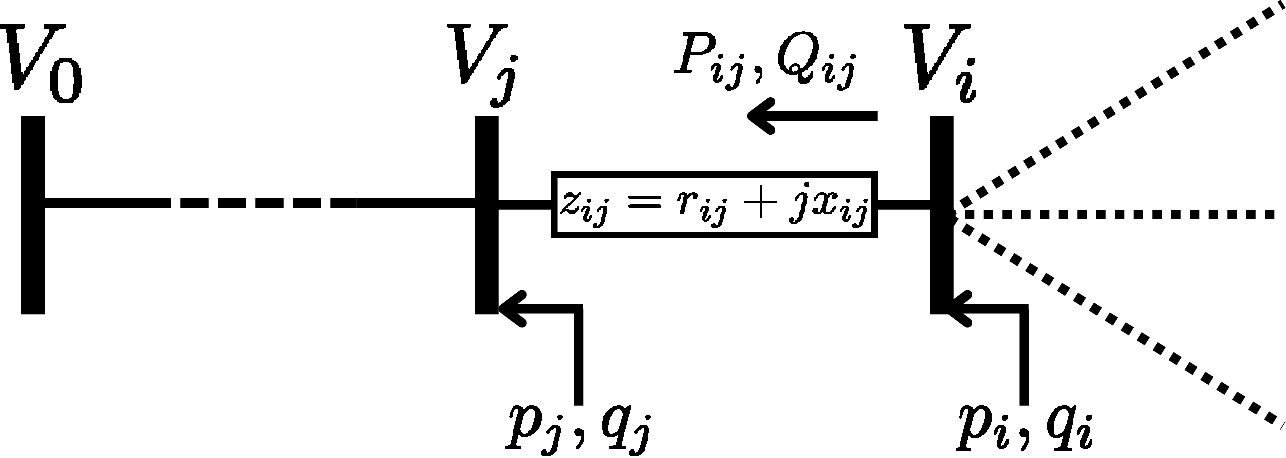}
    \caption{Representative radial network and notation.}
    \label{fig:radial_network}
\end{figure}

Using radial network shown in Fig.~\ref{fig:radial_network}, the \textit{DistFlow} equations can be represented as:
\begin{subequations}
\label{DistFlow_eqs}
\begin{align}
    P_{ij} &= p_{i} + \sum_{h:h \to i}(P_{hi} - r_{hi}l_{hi}), \\
    Q_{ij} &= q_{i} + \sum_{h:h \to i}(Q_{hi} - x_{hi}l_{hi}), \\
    v_i &= v_j + 2(r_{ij}P_{ij} + x_{ij}Q_{ij}) -|z_{ij}|^2l_{ij}, \\
    l_{ij} &= \frac{P_{ij}^2 + Q_{ij}^2}{v_i}, \label{eq: branch_current_eq}
\end{align}
\end{subequations}
where the net real power at node~$i \in \mathcal{N}$ is given by~$p_i=p_{g,i}-p_{d,i}$ where~$ p_{g,i} $ is the active power generation and~$p_{d,i}$ is the active power demand. Similarly, the net reactive power at node~$i$ is~$q_i=q_{g,i}-q_{d,i}$, where~$q_{g,i}$ is the reactive power generation and $ q_{d,i}$ is the reactive power demand. $\sum_{h:h \to i}(P_{hi} - r_{hi}l_{hi})$ indicates that the summation is over all nodes $h$ that send power to node $i$.

Applying the compact matrix formulation from~\cite{heidari2017}, for a distribution network with $N$ nodes, the vectors are defined as: active power flow from bus $i$ to $j$, $P := [P_{ij}]_{(i, j) \in \mathcal{L}} \in \mathbb{R}^{N}$, reactive power flow from bus $i$ to $j$, $Q := [Q_{ij}]_{(i, j) \in \mathcal{L}} \in \mathbb{R}^{N}$, the real component of demand, $p_d := [p_{d,i}]_{i \in \mathcal{N}} \in \mathbb{R}^{N}$, the imaginary component of demand $q_d:= [q_{d,i}]_{i \in \mathcal{N}} \in \mathbb{R}^{N}$, active and reactive power injections into bus $i$, $p:= [p_i]_{i \in \mathcal{N}} \in \mathbb{R}^{N} $, $q:= [q_i]_{i \in \mathcal{N}} \in \mathbb{R}^{N}$, active and reactive power generation $p_g:= [p_{g,i}]_{i \in \mathcal{N}} \in \mathbb{R}^{N} $, $q_g:= [q_{g,i}]_{i \in \mathcal{N}} \in \mathbb{R}^{N}$, squared branch current magnitudes on branch ($i,j$), $l:= [l_{ij}]_{(i,j) \in \mathcal{L}} \in \mathbb{R}^{N}$, and the squared voltage at each bus $i$ is $V := [v_i]_{i \in \mathcal{N}} \in \mathbb{R}^{N}$. The head node voltage is $V_0 := v_0\bold{1}_N \in \mathbb{R}^{N}$. Denote the matrices as the resistance matrix~$R := \text{diag} \{ r_{ij} \}_{(i,j) \in \mathcal{L}} \in \mathbb{R}^{N \times N}$, the reactance matrix $X := \text{diag} \{ x_{ij} \}_{(i,j) \in \mathcal{L}} \in \mathbb{R}^{N \times N}$, the impedance matrix $Z^2 := \text{diag} \{ z^2_{ij} \}_{(i,j) \in \mathcal{L}} \in \mathbb{R}^{N \times N}$. $I_N \in \mathbb{R}^{N \times N}$ being the identity matrix and $0_N \in \mathbb{R}^N$, the matrices that describe the network topology and impedance parameters are $A := [0_N \quad I_N] E - I_N$, $C := (I_N - A)^{-1}$, $D_R := CAR$, $D_X := CAX$, $M_p := 2C^\top RC$, $M_q := 2C^\top XC $ and $H := C^\top (2(RD_R + XD_X) + Z^2)$. Thus, the compact matrix expressions can be represented as:
\begin{subequations}
\label{eq:compact_PQVl}
\begin{align}
    P &= Cp - D_Rl, \label{eq:P_equation}\\
    Q &= Cq - D_Xl, \label{eq:Q_equation}\\
    V &= V_0 + M_p p + M_q q - Hl, \label{eq:V_equation}\\
    l &= \text{diag}\{V\}^{-1}(\text{diag}\{P\}P + \text{diag}\{Q\}Q) \label{eq:l_equation}.
\end{align}
\end{subequations}

To determine the hosting capacity of the grid, an optimization problem can be formulated that maximizes the total active power injections subject to the grid constraints defined by the system model equations in~\eqref{eq:compact_PQVl}. Despite~\eqref{eq:P_equation},~\eqref{eq:Q_equation},~\eqref{eq:V_equation} representing linear relationships, the presence of the nonlinear equation~\eqref{eq:l_equation} makes the optimization problem nonconvex. Convex relaxations and approximations are effective tools for addressing nonconvex optimization problems to obtain tractable formulations; however, their exactness is only guaranteed under some conditions and might not exactly satisfy the actual power flow equations~\cite{CvxRelax_SOCP, molzahn_survey_2019}. In this study, we employ convex inner approximations to address the nonconvexity of the AC OPF solution set, ensuring that voltage levels remain within operational limits. The following section motivates the need for this inner approximation through a small example.

\subsection{Motivating Example}\label{sec:motivating_example}
\begin{figure}[tbp]
    \centering
\includegraphics[width=0.7\linewidth]{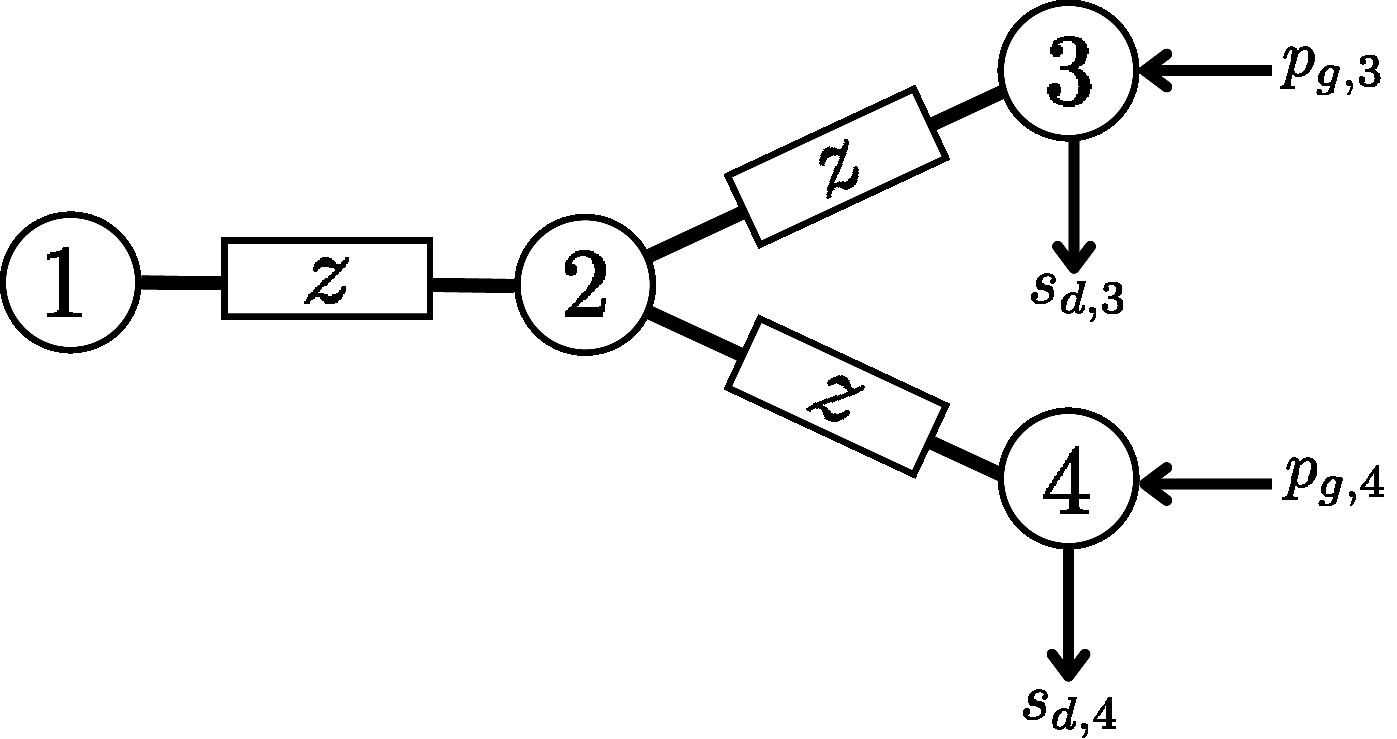}
    \caption{4-bus radial network with power injections at buses 3 and 4.}
    \label{fig:4bus_network}
\end{figure}

Consider the 4-bus radial network example shown in Fig.~\ref{fig:4bus_network}. The system is modeled with a base power of $S_\text{base} = 100$ MVA and a base voltage of $V_\text{base} = 4.16$ kV. Node 3 has a load of $s_{d,3} = -0.02 + i0.005$ pu and a power injection of $p_{g,3}$, while node 4 has a load of $s_{d,4} = -0.015 + j0.01$ pu and a power injection of $p_{g,4}$, with all lines having an impedance of $z = 0.55 + j1.33$ pu. $p_{g,3}$ and $p_{g,4}$ are controllable power injections. Using MatPower~\cite{zimmerman2011matpower}, we sweep $p_g$ and check whether the resulting voltages at each node fall within the acceptable limits of $[0.95, 1.05]$. If they do, we save the values to characterize the admissible region, which is defined as follows: 

\begin{definition}
\label{def:admissible_set}
\textbf{(AC Admissibility)} The set of power injections that satisfy the grid voltage and branch current limits (i.e., $v_j \in [\overline{v_j}, \underline{v_j}] \forall j \in \mathcal{N} \text{ and } l_{ij} \in [\overline{l_{ij}}, \underline{l_{ij}}] \forall \{i,j\} \in \mathcal{L})$ is referred to as the AC admissible set.
\end{definition}

\begin{figure}[tbp]
    \centering
    \includegraphics[width=0.8\linewidth]{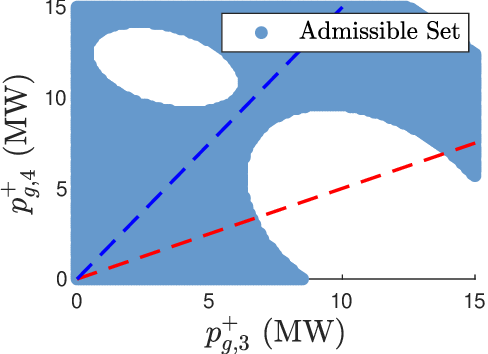}
    \vspace{0.5cm}
    \includegraphics[width=0.45\linewidth]{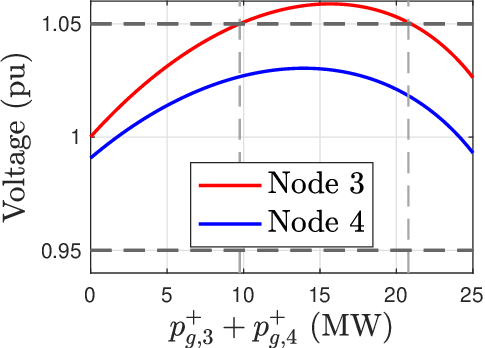}
    \hfill
    \includegraphics[width=0.45\linewidth]{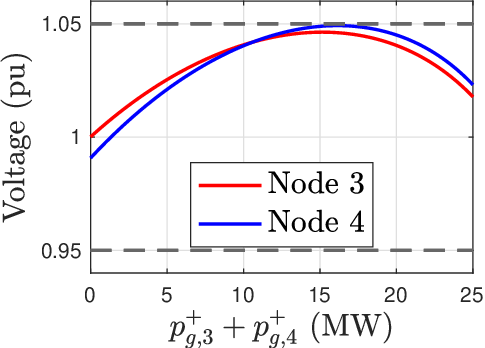}
    \caption{(Top) Admissible set of power injections with admissible (dark blue) and inadmissible (red) trajectories. Voltages at nodes 3 and 4: (Bottom Left) red trajectory passing through a hole (Bottom Right) vs. blue trajectory without hole passage.}
    \label{fig:motiv_ex}
\end{figure}

The admissible set, illustrated in Fig.~\ref{fig:motiv_ex} (Top) is nonconvex. Holes in the admissible set result from voltage violations, such as the rightmost hole caused by Node 3 voltages exceeding the 1.05 pu limit as power injections increase. In Fig.~\ref{fig:motiv_ex}, the red trajectory shows these voltage violations, while the blue trajectory remains within limits but shows a voltage drop beyond a certain injection level.
Selecting points beyond the holes in the admissible set, e.g., $(p_{g,3},p_{g,4}) = (12,12)$MW might result in a larger HC, but it compromises the ability to guarantee reliable operation, because, if $p_{g,3}$ or $p_{g,4}$ decreases, then the voltage can increase enough to land in one of the ``holes'' an overload. To address this, we restrict the admissible set to a convex inner approximation or restriction. A CIA offers a reliable method to approximate the admissible set from within, ensuring that all points in the set correspond to safe and reliable operating conditions. Formulating the inner approximation requires setting up a convex optimization problem that follows directly from~\cite{NAWAF2021}, which will be explained in the next section.


\subsection{Convex Inner Approximation Preliminaries} \label{subsec:convex_problem}
 The function~\eqref{eq:l_equation} is convex with respect to $P$, $Q$, and $V$ as its Hessian is positive semi-definite, but the set is nonconvex due to its non-linearity. To address this non-linearity introduced by~\eqref{eq:l_equation}, we can bound the function by using convex envelopes so that we can define upper and lower bounds, referred to as proxy variables, to eliminate the nonconvexity in the set caused by~\eqref{eq: branch_current_eq}. Thus, using the convex representations, a convex inner approximation can be formed.

The envelopes are constructed by introducing proxy variables, with the upper bound represented by $l^+$ and the lower bound by $l^-$, defined as:
\begin{align}
\label{convex_envelope}
    l^- \leq l(P, Q, V) \leq l^+.
\end{align}
Without knowing the exact values of $l(P, Q, V)$, it is guaranteed that as long as the proxy variables satisfy their own limits, the actual $l_{ij} \forall \{i, j\} \in \mathcal{L}$  will also remain within their limits.

The function $l$ is implicitly dependent on $P$, $Q$, and $V$. Therefore, when introducing proxy variables for $l$ based on the relationships defined by $P$, $Q$, and $V$, we must also define corresponding proxy variables for their upper and lower bounds as follows:
\begin{subequations}
\label{proxy_variables}
\begin{align}
\label{upper_proxy_P}
    P^+ &:= Cp - D_Rl^-, \\
    P^- &:= Cp - D_Rl^+, \\
    Q^+ &:= Cq - D_X^+l^- - D_X^-l^+, \\
    Q^- &:= Cq - D_X^+l^+ - D_X^-l^-, \\
    V^+ &:= V_0 + M_p p + M_q q - H^+l^- - H^-l^+, \\
\label{lower_proxy_V}
    V^- &:= V_0 + M_p p + M_q q - H^+l^+ - H^-l^-,
\end{align}
\end{subequations}
where $H^+$ and $D_X^+$ include non-negative elements and $H^-$ and $D_X^-$ include negative elements ($D_X = D_X^+ + D_X^-, \quad H = H_X^+ + H_X^-$).

To formulate the bounds around non-linear~\eqref{eq: branch_current_eq}, we first apply Taylor's Theorem around the nominal operating point $x_{ij}^0:= [P_{ij}^0 \quad Q_{ij}^0 \quad  v_i^0]^\top$, that is computed via running a load flow in PowerModels~\cite{powerModels}:
    \begin{equation}
        l_{ij} \approx  l_{ij}^0 + \mathbf{J}_{ij}^\top \delta_{ij} + \frac{1}{2} \delta_{ij}^\top \mathbf{H}_{e,ij} \delta_{ij},
    \end{equation}
   where $l_{ij}^0$ is the nominal branch current computed with the load flow simulation. The Hessian $\mathbf{H}_{e} \in \mathbb{R}^{n \times 3 \times 3}$, Jacobian $\mathbf{J}_{ij} := \text{col}\{\frac{\partial l_{ij}}{\partial P_{ij}}, \frac{\partial l_{ij}}{\partial Q_{ij}}, \frac{\partial l_{ij}}{\partial v_{i}}\}_ {(i,j) \in \mathcal{L}}$, and $\delta_{ij}$ are defined as,
\begin{equation}
\label{jacobian}
    \mathbf{J}_{ij} := 
    \left[
    \begin{array}{c}
    \frac{2P_{ij}^0}{v_{i}^0} \\
    \frac{2Q_{ij}^0}{v_{i}^0} \\
    -\frac{(P_{ij}^0)^2+(Q_{ij}^0)^2}{(v_{i}^0)^2}
    \end{array}
    \right]^\top, 
    \begin{aligned}
    \delta_{ij} &:= \begin{bmatrix}
    P_{ij}^\bullet - P_{ij}^0  \\
    Q_{ij}^\bullet - Q_{ij}^0 \\
    v_{i}^\bullet - v_{i}^0
    \end{bmatrix},
\end{aligned}
\end{equation}
\begin{equation}
        \mathbf{H}_{e,ij} := 
        \begin{bmatrix}
        \frac{2}{v_i^0} & 0 & \frac{-2P_{ij}^0}{(v_i^0)^2} \\
        0 & \frac{2}{v_i^0} & \frac{-2Q_{ij}^0}{(v_i^0)^2} \\
        \frac{-2P_{ij}^0}{(v_i^0)^2} & \frac{-2Q_{ij}^0}{(v_i^0)^2} & 2\frac{(P_{ij}^0)^2 + (Q_{ij}^0)^2}{(v_i^0)^3}
        \end{bmatrix},
\end{equation}
where $P_{ij}^\bullet$, $Q_{ij}^\bullet$, and $v_{i}^\bullet$ represent any of the proxies $ P_{ij}^+ $, $ P_{ij}^- $, $ Q_{ij}^+ $, $ Q_{ij}^- $, $ v_i^+ $, or $ v_i^- $. The Hessian is PSD with non-negative eigenvalues, indicating that $l_{ij}$ is a convex function. To derive the lower bound for branch currents, we use the first-order approximation of the function, which provides a lower bound (a global underestimator) for the convex function~\cite{boyd2004}. Thus, we can determine a lower proxy variable to establish a lower bound for $l_{ij}$ as follows:
\begin{equation}
\label{eq:lower_bound}
l_{ij}^- := l_{ij}^0 + (\mathbf{J}_{ij}^+)^\top \delta_{ij}^{-} + (\mathbf{J}_{ij}^-)^\top \delta_{ij}^{+} \leq l_{ij},
\end{equation}
where the Jacobian is separated into its positive and negative elements as $\mathbf{J}_{ij} := \mathbf{J}_{ij}^+ + \mathbf{J}_{ij}^-$. $\delta_{ij}^{+} $ is a function of the upper proxy variables $P_{ij}^+$, $Q_{ij}^+$, and $v_{i}^+$, while $\delta_{ij}^{-}$ is a function of the lower proxy variables $P_{ij}^-$, $Q_{ij}^-$, and $v_{i}^-$.

The prior study~\cite{NAWAF2021} utilizes a conservative upper bound formulation, which may limit the accuracy and increase computational burden as it will need more iterations to find the biggest possible HC limits of the grid within a single timestep. Those limits are engendered as:
\begin{subequations}
    \begin{align}
    l_{ij} &= |l_{ij}| \approx \left| l_{ij}^0 + \mathbf{J}_{ij}^\top \delta_{ij} + \frac{1}{2} \delta_{ij}^\top \mathbf{H}_{e,ij} \delta_{ij} \right|,\\
    &\leq \left| l_{ij}^0 \right| + \left| \mathbf{J}_{ij}^\top \delta_{ij} \right| + \left| \frac{1}{2} \delta_{ij}^\top \mathbf{H}_{e,ij} \delta_{ij} \right|,\\
    &\leq l_{ij}^0 + \max \left\{ 2 \left| \mathbf{J}_{ij}^\top \delta_{ij} \right|, \left| \delta_{ij}^\top \mathbf{H}_{e,ij} \delta_{ij} \right| \right\},\\
    &l_{ij} \leq l_{ij}^0 + \max \left\{ 2 \left| \mathbf{J}_{ij}^+ \delta_{ij}^+ + \mathbf{J}_{ij}^- \delta_{ij}^- \right|, \psi_{ij} \right\} \leq l_{ij, \text{from~\cite{NAWAF2021}}}^+,
\end{align}
\end{subequations}
where $\psi_{ij} := \max \left\{ (\delta_{ij}^{+,-})^\top \mathbf{H}_{e,ij} (\delta_{ij}^{+,-}) \right\}$ represents the maximum of eight possible combinations using the proxy variables of $P_{ij},Q_{ij},v_i$.

\subsection{Improvement of the Inner Approximation Bounds}
\label{subsec:bounds_improvement}
To build on~\cite{NAWAF2021}, we provide a less conservative bound (and a tighter convex envelope) to improve the approximation of the original non-linear function. The new upper bound $l_{ij, \text{SOC}}^+$ is obtained from the epigraph relaxation of~\eqref{eq: branch_current_eq} as a second-order cone (SOC) constraint:

\begin{equation} \label{eq:SOC}
\begin{split}
\left\lVert
        \begin{bmatrix}
        2P_{ij}^\bullet\\
        2Q_{ij}^\bullet \\
        l_{ij, \text{SOC}}^+ - v_i^-
        \end{bmatrix}
        \right\rVert_2 \leq l_{ij, \text{SOC}}^+ + v_i^-,\\
 P_{ij}^\bullet \in \{P_{ij}^+, P_{ij}^-\}, \quad Q_{ij}^\bullet \in \{Q_{ij}^+, Q_{ij}^-\}.
\end{split}
\end{equation}
In \eqref{eq:SOC}, the terms $P_{ij}^\bullet$ and $ Q_{ij}^\bullet$  represent proxy components that can be either positive or negative. This results in four possible combinations: $ (P_{ij}^+, Q_{ij}^+) $, $ (P_{ij}^-, Q_{ij}^-) $, $ (P_{ij}^+, Q_{ij}^-) $, and $ (P_{ij}^-, Q_{ij}^+) $. Since $ V^+(p) \ge V(p) \ge V^-(p) > 0 $ for all $p$, we can use  $V^-(p)$ in the denominator to engender the upper bound for~\eqref{eq: branch_current_eq}. 

\begin{figure}[t]
    \centering
\includegraphics[width=0.9\linewidth]{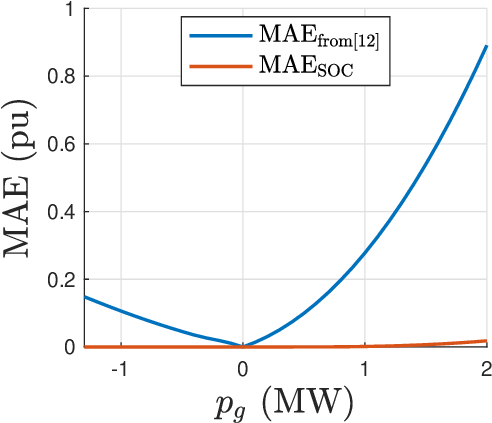}
    \caption{Comparing the MAE between~\eqref{eq: branch_current_eq} and the conservative upper bound $l_{ij,\text{from~\cite{NAWAF2021}}}^+$ (blue), and MAE between~\eqref{eq: branch_current_eq} and the proposed upper proxy variable $l_{ij, \text{SOC}}^+$ (red) across all $l_{ij} \in \mathcal{L}$. }
    \label{fig:comparison}
\end{figure}

\begin{figure}[t]
    \centering  
    \includegraphics[width=0.9\linewidth]{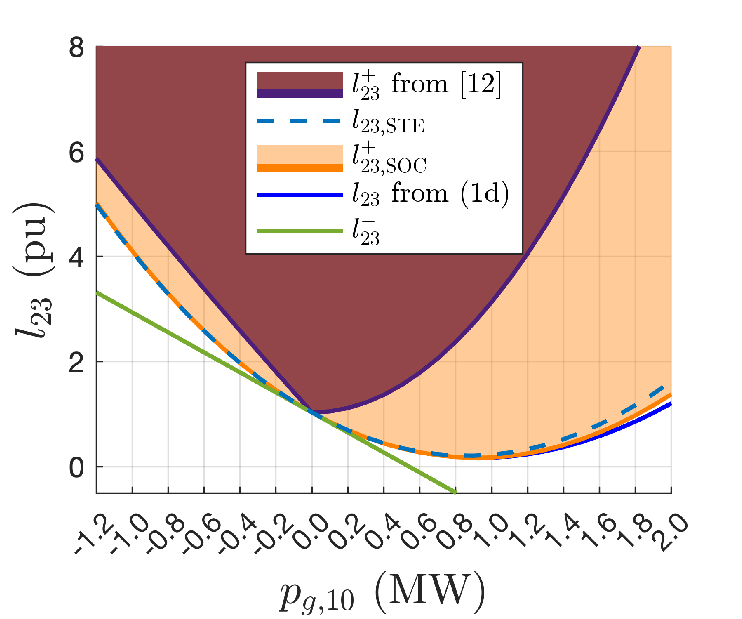} 
    \caption{Comparison of the branch currents: the upper bound proxy variable from~\cite{NAWAF2021} ($l_{ij}^+ \text{ from~\cite{NAWAF2021}}$), the second order Taylor expansion (STE) of~\eqref{eq: branch_current_eq} ($l_{ij, \text{STE}}$), the proposed upper bound ($l_{ij, \text{SOC}}^+$) obtained via epigraph relaxation using a SOC formulation, the actual branch current ($l_{ij}$) from~\eqref{eq: branch_current_eq}, and the lower bound proxy ($l_{ij}^-$) from~\eqref{eq:lower_bound}.
    }
    \label{fig:branch_currents}
\end{figure}

In Fig.~\ref{fig:comparison}, we compare the mean absolute error (MAE) between the actual branch current~\eqref{eq: branch_current_eq} and the proposed upper proxy variable $l_{ij, \text{SOC}}^+$ as well as between the actual branch current and the more conservative upper bound $l_{ij, \text{from~\cite{NAWAF2021}}}^+$ using,
\begin{equation}
   \text{MAE}_l= \frac{1}{L} \sum_{(i,j) \in \mathcal{L}} \left| l_{ij} - l_{ij}^+ \right|,
\end{equation}
where $L$ is the number of branches in the network.
 For all $p_g$ ranging from $[-1.2, 2]$ MW, injected at node 10 in IEEE37 network shown in Fig.~\ref{fig:36bus_feeder} (Left), all branch currents are computed. The MAE between~\eqref{eq: branch_current_eq} and the proposed $l_{ij, \text{SOC}}^+$ are significantly smaller compared to the conservative bound from \cite{NAWAF2021}.  As expected, the proposed bound satisfies $l_{ij, \text{SOC}}^+ \leq l_{ij, \text{from~\cite{NAWAF2021}}}^+$ for all $l_{ij} \in \mathcal{L}$. The differences between the actual branch current and the conservative upper bound $l_{ij, \text{from~\cite{NAWAF2021}}}^+$ ranging from $\approx [0, 0.89]$ pu. In comparison, the average difference between the actual branch current~\eqref{eq: branch_current_eq} and the proposed $ l_{ij, \text{SOC}}^+ $ across all branches in the network lies within a much tighter range, less than 0.01 pu.  Fig.~\ref{fig:branch_currents} illustrates the bounds around~\eqref{eq: branch_current_eq}, for branch $l_{23}$ in the IEEE37 network when injections are made at node 10. 

\subsubsection{Final Convex Problem Formulation}
Using the convex envelopes, the nodal generation limits can be determined by solving the convex optimization problem given by,
\begin{subequations} \label{P1_all}
\begin{align}
\textbf{(P1)} \quad \max_{p_{g}} \quad & f_0(p_g) \label{P1_obj} \\
\text{s.t.} \quad 
& P^+ = Cp - D_R l^- \label{P1_a} \\
& P^- = Cp - D_R l^+ \label{P1_b} \\
& Q^+ = Cq - D_X^+ l^- - D_X^- l^+ \label{P1_c} \\
& Q^- = Cq - D_X^+ l^+ - D_X^- l^- \label{P1_d} \\
& V^+ = V_0 + M_p p + M_q q - H^+ l^- - H^- l^+ \label{P1_e} \\
& V^- = V_0 + M_p p + M_q q - H^+ l^+ - H^- l^- \label{P1_f} \\
& l_{ij}^- = l_{ij}^0 + (\mathbf{J}_{ij}^+)^\top \delta_{ij}^- + (\mathbf{J}_{ij}^-)^\top \delta_{ij}^+  \label{P1_h} \\
& \left\lVert
    \begin{bmatrix}
    2P_{ij}^\bullet \\
    2Q_{ij}^\bullet \\
    l_{ij}^+ - v_i^-
    \end{bmatrix}
    \right\rVert_2 \leq l_{ij}^+ + v_i^-, \quad \bullet \in \{-, +\} \label{P1_g} \\
& p = p_g - p_d, \quad q = q_g - q_d \label{P1_i} \\
& P^+ \le \overline{P},\quad  Q^+ \le \overline{Q},\quad V^+ \le \overline{V} \label{P1_j} \\
&\underline{P}\le P^-,\quad  \underline{Q}\le Q^-, \quad   \underline{V}\le V^-. \label{P1_k}
\end{align}
\end{subequations}
where the objective function $ f_0(p_g) $ to be maximized can take different forms. For instance, $ \sum_{i=1}^N \alpha_i p_{g,i} $ or $ \sum_{i=1}^N \alpha_i \log(p_{g,i}) $, where the weight factor $\alpha_i$  could either be 1 (uniform weights) or $ \frac{p_{d,i}}{\sum_{i=1}^N p_{d,i}} $ (proportional to demand). The proxy variables are defined by the limits in equations~\eqref{P1_j} and~\eqref{P1_k}, which bound the allowable range for the proxies. Note that the reactive power injections to all nodes are $q_{g,i} := 0$ (unity power factor). 

\subsection{Increasing the Grid Hosting Capacity}
\begin{figure}[t]
    \centering
\includegraphics[width=0.8\linewidth]{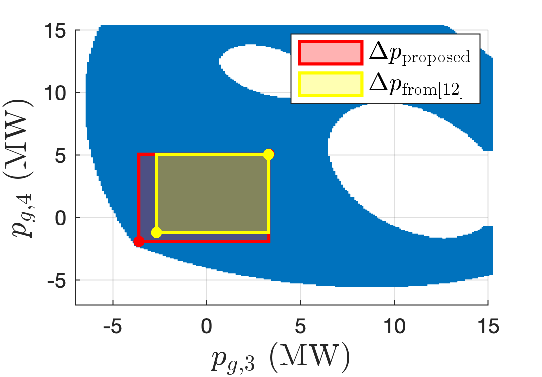}
        \caption{Impact of the proposed upper bound formulation of current on the results, showing a larger hyperrectangle compared to using the formulation in~\cite{NAWAF2021}.}
    \label{fig:hyperrectangle}
\end{figure}

By solving the optimization problem (\textbf{P1}), our goal is to determine the largest hyperrectangle (i.e., orthotope) $\Delta p_g$ (which defines the HC limits) that can fit in the nonconvex set of admissible injections. To determine the boundaries of this rectangle, we solve (\textbf{P1}) twice: first to find the upper limits ($p_g^+$), where $p_g \geq 0$, and then to find the lower limits ($p_g^-$), where $p_g \leq 0$. In both cases, we maximize and minimize the objective function $\sum_{i=1}^N \alpha_i p_{g,i}$, respectively. This allows us to define the boundaries of $\Delta p_g$ as $\{p~\mid~p_{g,i}^-~\leq~p~\leq~p_{g,i}^+, \forall i \in N\}$. Consequently, for a network with $N$ nodes, an $N$-dimensional hyperrectangle will be formed. Using the hyperrectangle approach ensures that the operation of DERs is independent, preventing any coupling between them that could lead to voltage violations. Hyperrectangle concept is illustrated with a two-dimensional example in Fig.~\ref{fig:hyperrectangle} using the 4-bus motivating example in Fig~\ref{fig:4bus_network}.
As shown in Fig.~\ref{fig:hyperrectangle}, the proposed convex bounds for branch currents in this study enlargen the volume of the hyperrectangle. This improvement is a result of updating the branch current bounds into tighter envelopes as explained in Section~\ref{subsec:convex_problem}. The proposed upper bound results in a larger hyperrectangle since $V^{+}_{\text{proposed}}(p) \leq V^{+}_{\text{from~\cite{NAWAF2021}}}(p)$, where $V^{+}_{\text{proposed}}(p) $ is the upper proxy for voltages as a function of the SOC upper bound $l_{ij, \text{SOC}}^+$ and $V^{+}_{\text{from~\cite{NAWAF2021}}}(p) $ is the upper proxy for voltages as a function of the upper bound proposed in~\cite{NAWAF2021}.

\begin{remark} \textbf{(Enlarging the Hyperrectangle)}
    In a radial network with \textit{N}~nodes with generation, the resulting hyperrectangle will be \textit{N-dimensional}. Therefore any improvement in one of the boundaries of the rectangle will significantly impact the volume of the hyperrectangle. Table~\ref{tab:example_bounds} compares the simulation results of the 4-bus motivational example and modified IEEE37 network in Fig.~\ref{fig:36bus_feeder} (Left). The difference between the results of using two methods for the 4-bus motivational example is shown in Fig.~\ref{fig:hyperrectangle}. In both the 4-bus example and the modified IEEE37 network, the proposed bounds represent a larger hyperrectangle compared to the old formulation in~\cite{NAWAF2021}.
\end{remark}
\begin{table}[tbp]
    \centering 
    \renewcommand{\arraystretch}{1.2} 
    \caption{Comparison of HC Limits after first iteration}
    \label{tab:example_bounds}
    \begin{tabular}{rcc}
        \toprule
        & \textbf{4-Bus} & \textbf{Modified IEEE37} \\ 
        \midrule
        \textbf{From~\cite{NAWAF2021}} & [ -3.89, 8.34] MW & [-2.64, 5.09] MW\\ 
        \textbf{Proposed} & [ -5.56, 8.34] MW& [-2.68, 5.09] MW\\ 
        \bottomrule
    \end{tabular}
\end{table}
While the proposed method improves solutions for a snapshot, the practical application of dynamic hosting capacity (DHC) requires consideration of time-varying demand. In real-world scenarios, such limits would need to be updated frequently, every 5–15 minutes, for instance, to reflect changes in demand. This naturally raises the question of how CIA can be effectively applied to address the challenges of time-varying demand. These aspects are further explored in the following section.

\section{Fairness in Hosting Capacity Allocation: Time \& Space}
\label{sec:Fairness_of_CIA}
\label{sec:fairness}
\begin{figure}[tbp]
    \centering
    \includegraphics[width=0.425\linewidth]{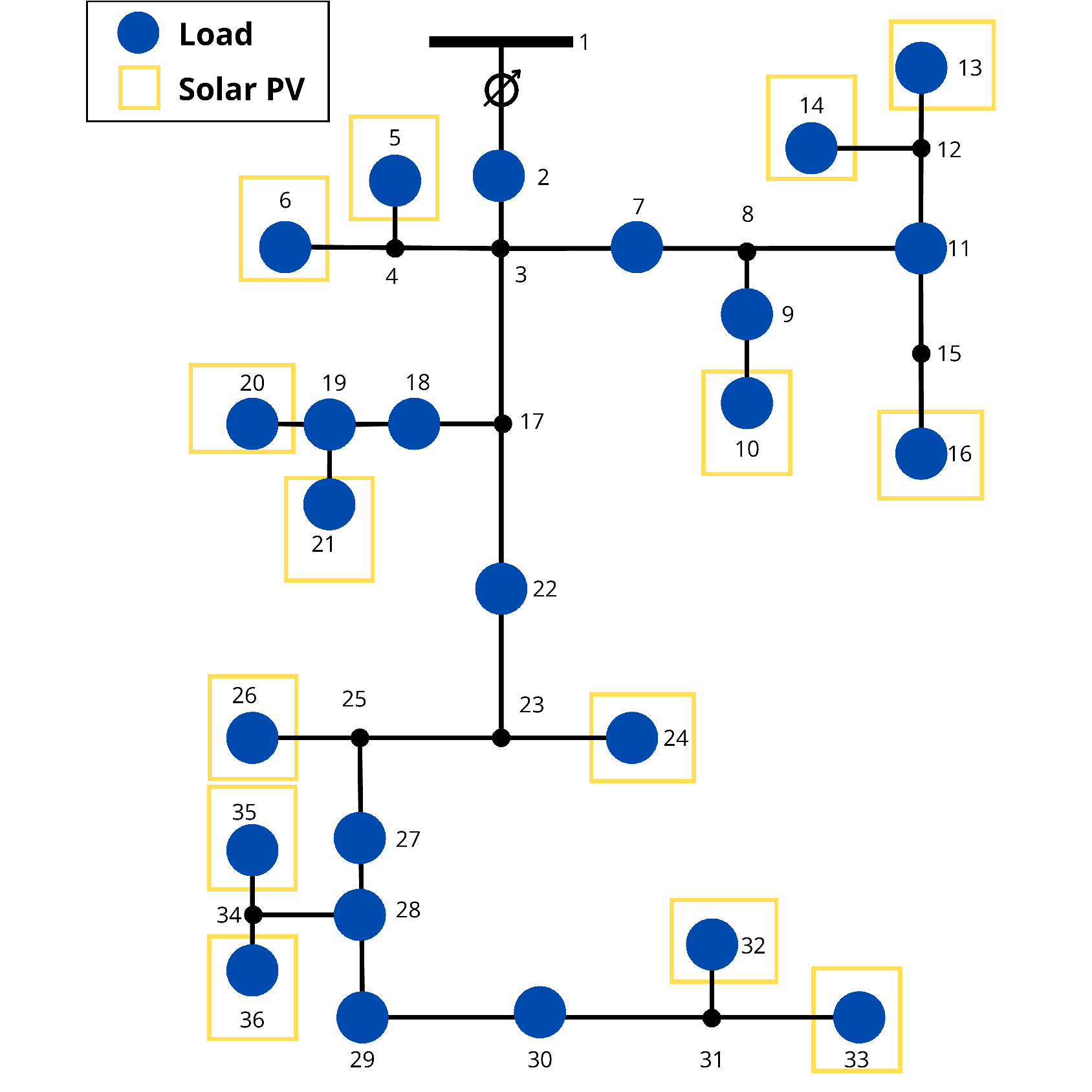}
    \hfill
    \includegraphics[width=0.54\linewidth]{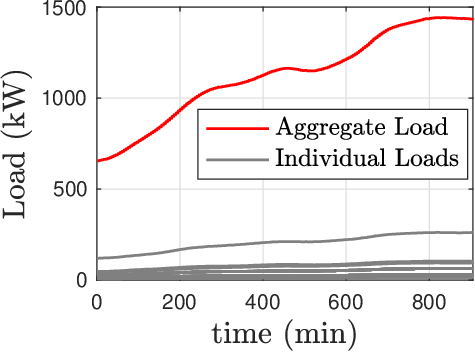}
    \caption{(Left) Modified IEEE-37 to 36-bus distribution feeder that consists of 14-DERs and 25 loads. (Right) Aggregate load profile (red) and individual node load profiles (grey) for the IEEE-37 system.}
    \label{fig:36bus_feeder}
\end{figure}

This section examines the unfairness that arises when solving (\textbf{P1}) to determine the HC of the grid using the time-varying demand data sourced from~\cite{almassalkhi2020hierarchical}. The load profile of the data used is given in Fig.~\ref{fig:36bus_feeder} (Right). Note that the data is limited to daytime hours, as these are the hours of interest in solar PV HC analysis which is presented in Section~\ref{sec:PV_Curtailment}.  Depending on the choice of objective function used while maximizing HC, the results can lead to inequitable distribution among nodes. To illustrate this, a case study is conducted using the 36-bus radial network, which is a modification of the IEEE-37 distribution feeder shown in Fig.~\ref{fig:36bus_feeder} (Left).

\subsection{Example of Inequity in Hosting Capacity Allocation}
\label{sec:inequities_in_HC}
\begin{figure}[t]
    \centering
    \includegraphics[width=0.6\linewidth]{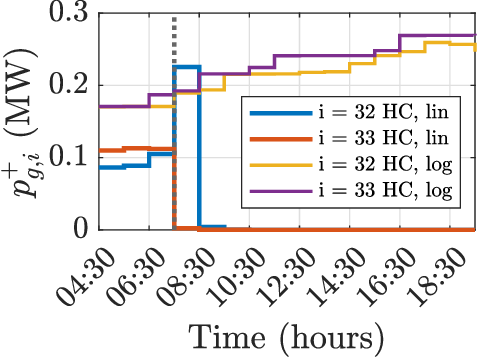}
    \caption{Hourly HC values for two nodes in the system, computed using logarithmic and linear objective functions.}
    \label{fig:unfair_HC_case}
\end{figure}

The optimization results may favor certain nodes, leading to a situation where one node's HC is significantly higher while others receive only a small share. This disparity primarily arises due to the nodes' locations within the network. For instance, nodes situated closer to the slack bus tend to dominate in terms of HC. We refer addressing this issue as spatial fairness. Fig.~\ref{fig:unfair_HC_case} illustrates a case where an unfair allocation occurs between nodes 32 and 33, when the objective is to maximize the total injections using linear objective (blue and red lines) in (\textbf{P1}). We see that at time $t=15$, the optimization prioritizes one node over another to maximize the HC. After some time, both nodes receive zero injections in favor of other nodes within the network. 

\begin{figure}[t]
    \centering
    \includegraphics[width=0.49\linewidth]{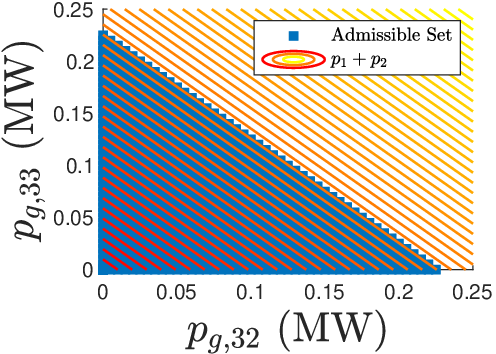}
    \hfill
    \includegraphics[width=0.49\linewidth]{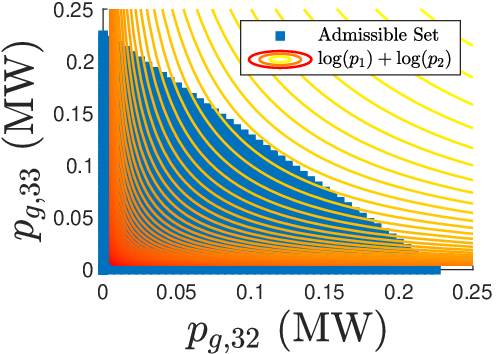}
    \caption{(Left) Admissible set with linear objective function $f(p_g) = \sum_{i = 1}^N p_{g,i}^+$. (Right) Admissible set with logarithmic objective function $f(p_g) = \sum_{i = 1}^N \log(p_{g,i}^+)$}
    \label{fig:unfair_allocation}
\end{figure}

The situation in Fig.~\ref{fig:unfair_HC_case} arises because the objective function is flat over the admissible set. At time $t=15$, when inequity occurs, we can observe that using the linear objective $ f_0(p_g) = \sum_{i = 1}^N p_{g,i}^+ $, represented by contour lines in Fig.~\ref{fig:unfair_allocation} (Left), results in multiple optimal solutions within the admissible set. As the objective achieves the same maximum value at various points, it often leads to selecting extreme points, resulting in inequitable outcomes. 

By changing the objective function from maximizing linear to logarithmic sum of injections as in Fig.~\ref{fig:unfair_allocation} (Right), the imbalance between nodal limits is prevented and spatially fairer results are obtained, as shown with purple and yellow lines in Fig.~\ref{fig:unfair_HC_case}. In this paper, fair allocation is defined proportional to each node's load, meaning that nodes with higher demand should receive a larger hosting capacity. To achieve this, we introduce a weighting factor $\alpha_i = \frac{p_{d,i}}{\sum_{i=1}^N p_{d,i}} \forall i\in \mathcal{N}$ in the objective function, allowing the allocation to reflect each node's demand.
 
\label{sec:methods}
With the weighting factor in the model, we introduce four different objective function designs, summarized as follows:
\begin{itemize}
    \item \textbf{Scenario~1:}~The unweighted linear objective, $f_0(p_g)~=~\alpha^\top p_{g}^+$, where $\alpha = \bold{1}_N$.
    \item\textbf{Scenario~2:}~The weighted linear objective, $f_0(p_g)~=~\alpha^\top p_{g}^+$, where $\alpha_i = \frac{p_{d,i}}{\sum_{i=1}^N p_{d,i}}, \forall i \in \mathcal{N}$.
    \item \textbf{Scenario~3:}~The logarithmic approach with the unweighted logarithmic objective, $f_0(p_g)~=~\alpha^\top \log(p_{g}^+)$, where $\alpha = \bold{1}_N$.
    \item \textbf{Scenario~4:}~The weighted logarithmic objective, $f_0(p_g)~=~\alpha^\top \log(p_{g}^+)$, where $\alpha_i~=~\frac{p_{d,i}}{\sum_{i=1}^N p_{d,i}}, \forall i \in \mathcal{N}$.
\end{itemize}

Scenario 1 maximizes the sum of injections without giving preference to any particular node. Scenario 2 assigns higher priority to nodes with larger demand by weighting injections proportionally to demand. To prevent any node from being eliminated, we use Scenarios 3 and 4, ensure all nodes receive some injection with and without considering demand weights.

\begin{figure}[t]
    \centering
    \includegraphics[width=0.8\linewidth]{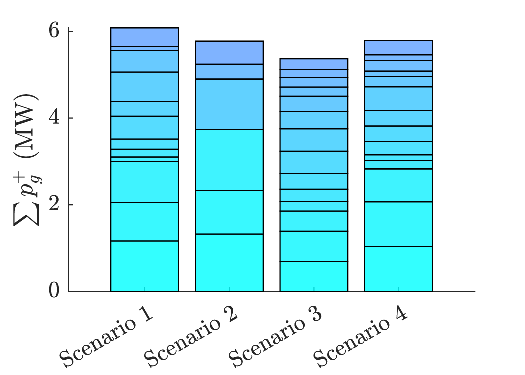}~
    \caption{Hosting capacity at time $t$ with four different objective functions.}
    \label{fig:HC_bar_chart}
\end{figure}

\begin{table}[t]
    \centering 
    \renewcommand{\arraystretch}{1.2} 
    \caption{Aggregate HC and Non-Zero Nodes Across Four Scenarios at Time $t$.}
    \label{tab:4Scenarios}
    \begin{tabular}{rcc}
        \toprule
        & \textbf{$\sum_{i=1}^{N}p_{g,i}^+$ (MW)} & \textbf{$\#$ of Nodes with Non-Zero HC} \\ 
        \midrule
        \textbf{Scenario 1} & 6.08 & 12 \\ 
        \textbf{Scenario 2} & 5.78 & 6 \\ 
        \textbf{Scenario 3} & 5.37 & 14\\ 
        \textbf{Scenario 4} & 5.79 & 14\\ 
        \bottomrule
    \end{tabular}
\end{table}
The bar chart in Fig.~\ref{fig:HC_bar_chart} illustrates the HC at a random time step $t$ for four of the scenarios given in Table~\ref{tab:4Scenarios}. 
We can clearly see the trade-off between ensuring equitable access, which may reduce the aggregate HC, and maximizing utilization across all nodes. 
Fairness decisions can also be incorporated into the optimization problem via adding a SOC constraint from~\cite{sundar2024parametricsecondorderconerepresentable} as explained in the following subsection.
\subsection{Fairness-Aware Hosting Capacity}
\label{sec:fairness_HC}
In~\cite{sundar2024parametricsecondorderconerepresentable}, the authors leverage the well-known L1-L2 norm inequality to integrate fairness decisions in the optimization problem. Derived using the equivalence of norms~\cite{trefethen1997numerical}, L1-L2 norm inequality is given as:
\begin{equation}
\label{eq:l1_l2_norms}
    \| \mathbf{p} \|_2 \leq \| \mathbf{p} \|_1 \leq \sqrt N \cdot \| \mathbf{p} \|_2,
\end{equation}
where $\mathbf{p} \in \mathbb{R}^{N}_{\geq 0}$ is the vector of utilized resources (i.e., the hosting capacity) and $N$ is the number of agents (i.e., nodes) with generation. The left-hand inequality is satisfied with equality when \textit{one agent is allocated the full capacity} (i.e., $\| \mathbf{p} \|_2 = \| \mathbf{p} \|_1 $). The right-hand inequality is satisfied with equality when \textit{every agent is allocated the same capacity}. Denoting these two cases as unfair and fair, respectively, we can consider the following convex representation:

\begin{equation}
    \label{eq:fairness_constraint}
    (1 - \varepsilon + \varepsilon \sqrt{N}) \| \mathbf{p} \|_2 \leq \| \mathbf{p} \|_1,
\end{equation}
where $\varepsilon \in [0,1]$ is a parameter that explicitly constrains ``at least $\varepsilon$-fair,'' such that when $\varepsilon~=~0$ we loosen constraint to allow any solution, including the ``unfair'' solution, while  $\varepsilon~=~1$ tightens the constraint and effectively enforces the ``fair'' solution.

Thus, we can include~\eqref{eq:fairness_constraint} and preserve the convexity of \textbf{(P1)} in~\eqref{P1_all}.

Note that if we want to consider fairness of the HC allocations relative to the demand at each node, we can modify~\eqref{eq:fairness_constraint} as follows:
\begin{equation}
    \label{eq:fairness_constraint_prop}
    \left(1 - \varepsilon + \varepsilon \sqrt{N}\right) \cdot  \sqrt{\sum_{i=1}^N\left ( \frac{p_i}{\alpha_i} \right)^2} \leq \sum_{i=1}^N \frac{p_i}{\alpha_i},
\end{equation}
where $\alpha_i :=p_{d,i}/\sum_{i=1}^N p_{d,i} \quad i \in \mathcal{N}$. For example, if $\varepsilon = 1$, the capacities will be directly proportional to the demand at each node. Of course, it is assumed that all nodes with PV generation has a positive demand to ensure $\alpha_i>0$.

Thus, we can now use~\eqref{eq:fairness_constraint} and~\eqref{eq:fairness_constraint_prop} to introduce two new scenarios as modifications of Scenario~1 to consider constrained spatial $\varepsilon$-fairness
\footnote{Enforcing a lower bound on temporal fairness in the optimization problem requires a multi-period formulation, which is considered outside the scope of this paper. We will assess temporal fairness in post-processing of results.}:
\begin{itemize}
    \item \textbf{Scenario~1-F1:}~The unweighted linear objective, $f_0(p_g)~=~\alpha^\top p_{g}^+$, where $\alpha = \bold{1}_N$ and inequality constraint~\eqref{eq:fairness_constraint} added where $\varepsilon \in [0,1]$.
    \item \textbf{Scenario~1-F2:}~The unweighted linear objective, $f_0(p_g)~=~\alpha^\top p_{g}^+$, where $\alpha = \bold{1}_N$ and inequality constraint~\eqref{eq:fairness_constraint_prop} added where $\varepsilon \in [0,1]$.
 \end{itemize}

\begin{figure}[t]
    \centering
    \includegraphics[width=0.49\linewidth]{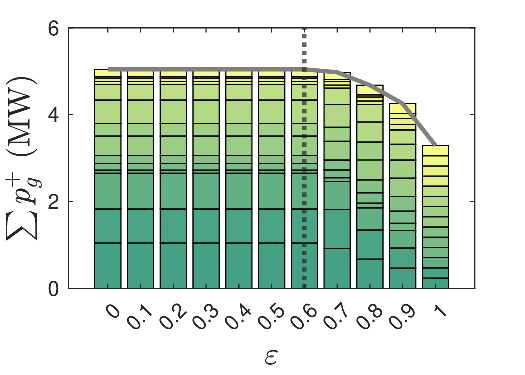}
    \hfill
    \includegraphics[width=0.49\linewidth]{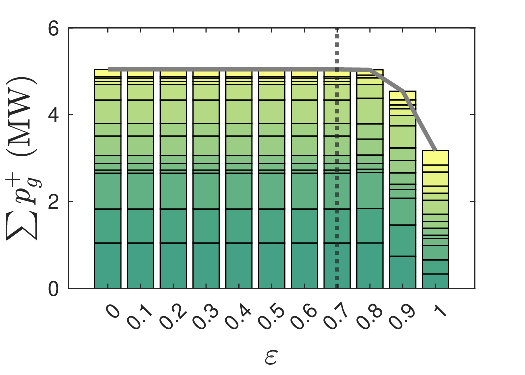}
    \caption{(Left) Impact of varying $\varepsilon$ on the HC at a snapshot on Scenario 1-F1. and (Right) on Scenario 1-F2.}
    \label{fig:epsilon_tradeoff}
\end{figure}

Note that adding  constraints~\eqref{eq:fairness_constraint} and~\eqref{eq:fairness_constraint_prop} in Scenarios~1-F will not impact infeasibility of the modified \textbf{(P1)}, since the solution satisfying $p_i=0~~\forall i=1,\hdots, N$ satisfies the $\varepsilon$-fairness constraints for any $\varepsilon \in [0,1]$ (i.e., is always feasible, if the original \textbf{(P1)} is feasible). Of course, there is a trade-off between maximizing the total HC and the degree to which the (spatial) distribution is fair. Thus, selecting an appropriate $\varepsilon$ is important to consider. To this end, we conduct simulation-based analysis on the IEEE37 system for different $\varepsilon$, which is shown in Fig.~\ref{fig:epsilon_tradeoff} for loading conditions representing the first time-step, per the dataset depicted in Fig.~\ref{fig:36bus_feeder}(Right). In Fig.~\ref{fig:epsilon_tradeoff}(Left), we show the effects of increasing $\varepsilon$. As $\varepsilon~\rightarrow~1$, \eqref{eq:fairness_constraint}, the total HC decreases once $\varepsilon \approx 0.6$, when the $\varepsilon$-fairness constraint becomes binding. When $\varepsilon=1$, the HC is distributed equally as expected in Scenario~1-F1. Similarly, for Scenario~1-F2, when $\varepsilon \approx 0.7$, \eqref{eq:fairness_constraint_prop} becomes binding and the HC values become relatively equal (i.e., in proportion with demand).

Now that we have discussed how the choice of $\varepsilon$ affects fairness in the allocation of HC, it is worth also relating the notion of $\varepsilon$-fairness to another common fairness metric: Jain's Fairness Index (JFI). JFI$:=\frac{||\mathbf{p}||_1^2}{N ||\mathbf{p}||_2^2}$ is a widely recognized measure for quantifying fairness in various resource allocation problems~\cite{jain1984quantitative}. For example, when resources $p_i$ are equal then JFI$=1$. As JFI decreases towards the worst-case ($1/N$), the $\mathbf{p}$ allocation becomes heavily skewed towards a few agents\footnote{If $\mathbf{p}=0$, then the JFI is undefined, which we define as JFI$=0$ herein.}. This is similar to the $\varepsilon$-fair constraint, which is related to JFI by considering the square of~\eqref{eq:fairness_constraint}, rearranging, and dividing both sides by $n$ to yield:
\begin{equation}
   \frac{(1 - \varepsilon + \varepsilon \sqrt{N})^2 }{N} \leq \frac{\| \textbf{p} \|_1^2}{N \cdot \| \textbf{p} \|_2^2}:=\textbf{JFI}.
\end{equation}
Thus, JFI is also bounded from below by the choice of $\varepsilon$. For example, $\varepsilon=0$ (or $\varepsilon=1$) means that the JFI is no smaller than $1/N$ (or $1$).

Beyond spatial fairness as described above, fairness over time is also a consideration, e.g., \textit{fair-over-time} from~\cite{moring}. Temporal fairness assesses whether there are significant changes in hosting capacity across time. Temporal fairness of HC highly depends on the time-varying nature of the demand, as fluctuations can lead to uneven hosting capacity over time. To better understand and evaluate the fairness of the presented methods, it is necessary to quantify fairness both spatially and temporally using a metric aligned with our definition of fairness, as explained in the next section. 

\subsection{Quantifying Fairness}
\label{sec:quantifying_fairness}
To evaluate the fairness of the designs of scenarios presented in Section~\ref{sec:methods} and~\ref{sec:fairness_HC} across both time and space, we leverage JFI. Fairness, as defined in our context, reflects a balanced distribution that aligns with the proportion of the demand each node have. Thus, define the ratio of nodal HC to the corresponding load as, 
\begin{equation}
\label{eq:prop_p_Pl}
    \rho_i(t) := \frac{p_{g,i}^+(t)}{p_{d, i}(t)},
\end{equation}
where $p_{g,i}^+(t)$ and $p_{d, i}(t)$ are the HC and load at time $t$ at node $i$, respectively. 
Ideally, the HC-to-demand ratio should be similar across all nodes, indicating equitable allocation. By incorporating this ratio into JFI, we quantify the fairness of HC allocation. A high JFI value over space reflects a fair allocation, where nodes with higher demand receive proportionally higher hosting capacities, aligning with the principle of proportional fairness.

\begin{figure}[t]
    \centering
    \includegraphics[width=0.49\linewidth]{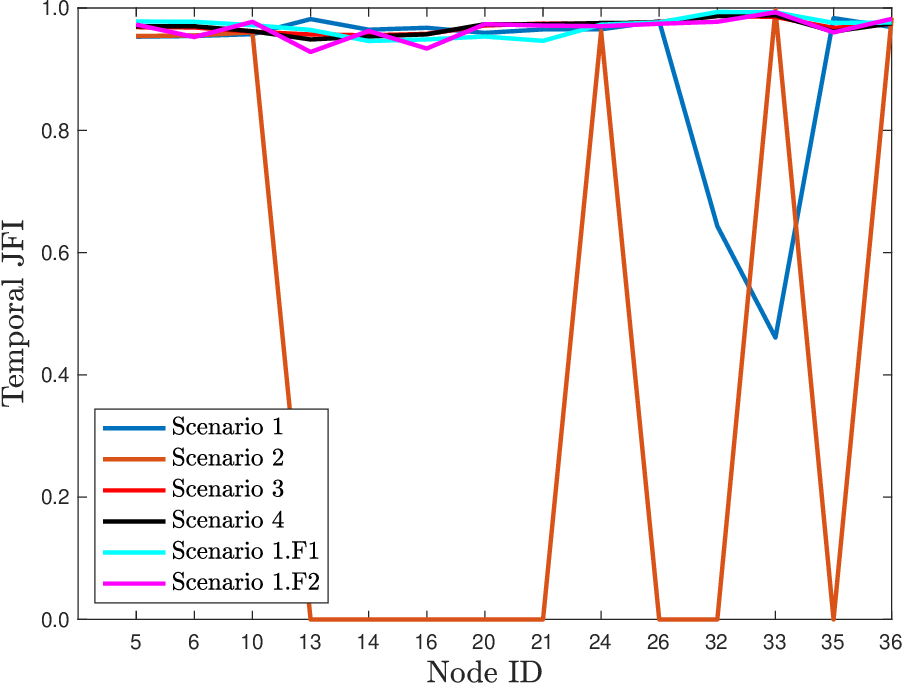}
    \hfill
    \includegraphics[width=0.49\linewidth]{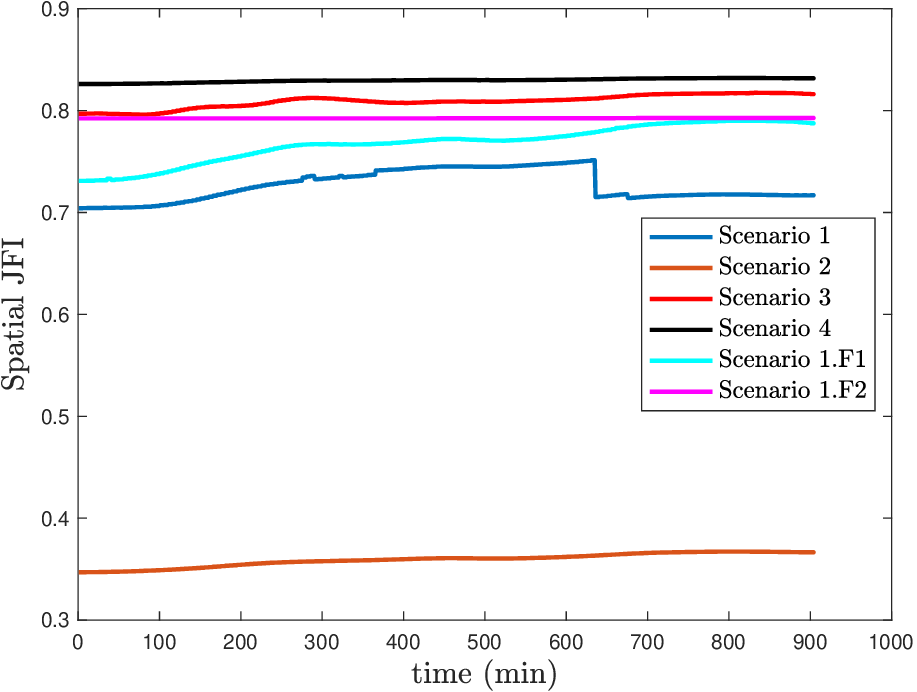}
    \caption{(Left) Temporal fairness for every node in the system, (Right) Spatial fairness for each time step of the six allocation methods.}
    \label{fig:combined_JFI}
\end{figure}

For each time step $t \in \mathcal{T}$, we can quantify the temporal and spatial fairness using JFI as,
\begin{subequations}
    \begin{align}
        \label{temporal_fairness}
    \mathcal{J}_{\text{temporal}}[i] = \frac{\left( \sum_{t=1}^T \rho_i(t) \right)^2}{T \cdot \sum_{t=1}^T \rho_i(t)^2 }, \quad i = 1, 2, \dots, N,\\
      \label{spatial_fairness}
\mathcal{J}_{\text{spatial}}(t) = \frac{\left( \sum_{i=1}^N \rho_i(t) \right)^2}{N \cdot \sum_{i=1}^N \rho_i(t)^2}, \quad t = 1, 2, \dots, T,
    \end{align}
\end{subequations}
where $\mathcal{J}_{\text{temporal}}[i]$ is the temporal fairness that denotes the fairness at node $n$ within its own HC over time and $\mathcal{J}_{\text{spatial}}(t)$ is the spatial fairness vector that denotes the fairness across nodes in each time step. Fig.~\ref{fig:combined_JFI} illustrates $\mathcal{J}_{\text{temporal}}[i]$ (Temporal JFI) and $\mathcal{J}_{\text{spatial}}(t)$ (Spatial JFI).

In Scenario 1's temporal fairness, we see a drop for nodes 32 and 33 due to the zero capacities assigned as explained in Section~\ref{sec:inequities_in_HC}. Similarly, Scenario 2 worsens the inequity, leaving certain nodes without any capacity, which results in temporal fairness scores of zero.
Logarithmic objectives (Scenarios 3 and 4) play a key role in promoting equitable allocation of hosting capacity, aiming to ensure that all nodes have access to some capacity and eliminating the unfair effects of Scenarios 1 and 2. The weighting factor ($\alpha$) provides flexibility in prioritizing capacity allocation based on specific needs, such as demand, inverter capacity, or other factors. This adaptability allows for designing objectives tailored to particular fairness criteria. 

In this study, we consider HC to be proportional to the demand at each node, which aligns with the definition of fairness in~\eqref{eq:fairness_constraint_prop}. In this example, Scenario 1-F2 is used with $\varepsilon = 0.85$, to ensure spatial fairness. 

\section{Case Study}
\label{sec:PV_Curtailment}
\subsection{Case Setup}
This case study uses the modified IEEE 37-bus system, shown in Fig.~\ref{fig:36bus_feeder}(Left) which illustrates the network topology, DER locations, and load points. Only leaf nodes are selected for the HC allocation problem to reflect a practical residential PV use case.

\begin{figure}[t]
    \centering
\includegraphics[width=0.8\linewidth]{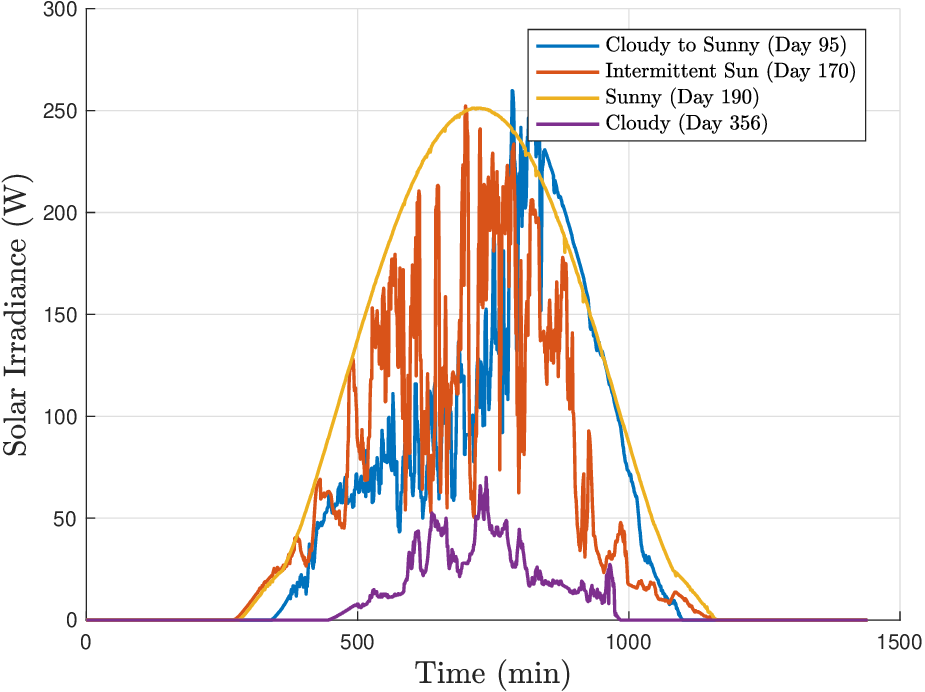}
    \caption{Representative days from minute-level solar irradiance data for a single panel used in the case study.}
    \label{fig:solar_data}
\end{figure}

The network consists of 36 buses and 35 branches. The demand profile is based on a realistic feeder demand dataset from~\cite{almassalkhi2020hierarchical}. We generate our demand by selecting single-phase demands from this feeder and assign to nodes in the IEEE-37 system. The generated aggregate demand profile of the network spans a range of 0.66 MW to 1.44 MW as shown in Fig.~\ref{fig:36bus_feeder} (Right).  We use real, minutely solar PV data from a 330~W solar panel over a full year. An example of data from the single solar panel for a few specific days is depicted in Fig.~\ref{fig:solar_data}. The solar PV panel data is scaled up to correspond to any larger PV array (e.g, 3.3kW arrays is modeled as ten times the 330W data).

Each node in the system has varying demand values, ranging from 7 kW to 63.8 kW, reflecting spatial heterogeneity, while the same solar generation profile is assumed across all nodes. Additionally, two alternative demand scenarios are considered (high and low demand, representing $\pm 25\%$ variations from the nominal demand shown in Fig.~\ref{fig:36bus_feeder} (Right)) to analyze how HC changes under different load conditions. Note that we compute HC at 5-minute intervals throughout the day, using demand profiles that are also captured at 5-minute resolution, to capture the dynamic interaction between demand and solar generation\footnote{For the purpose of calculating dynamic solar hosting capacity, the limits are computed only for daytime hours, excluding nighttime, since the analysis focuses solely on solar generation.}.

\subsection{Base-case solar PV capacity}
In this section, we analyze the results using the six scenarios~\footnote{The best choice of $\varepsilon$ is not straightforward due to its impact on the baseline, which is applied to each node and affects benefit and cost calculations. However, based on extensive simulations, selecting $\varepsilon=0.85$ results in the best performance for Scenarios 1-F2 when capacity increases are less than 60\%.} and demonstrate how a relatively small amount of annual solar curtailment (a few MWh) can significantly increase the overall PV capacity (many MWs). Traditionally the distribution system operator (DSO) determines the HC by considering a clear-sky day at noon, and it is assumed to coincide with the minimum aggregate household energy demand which results in a conservative HC. However, if curtailment is permitted and based on dynamic HC values, it is possible for the installed PV capacity to exceed the feeder's nominal PV hosting capacity.

A feeder's nominal (base-case) solar PV hosting capacity is determined by solving $\textbf{(P1)}$. In this study, the static limit is defined as the minimum DHC.
Since the DHC varies throughout the day and across nodes, the static limit is defined as
\begin{equation}
    L_{\text{pv},i} := \min_{t} (p_g^+(t)),
\end{equation}
where $L_{\text{pv},i}$ is the base-case hosting capacity of node $i$ (shown with red dots in Fig.~\ref{fig:sunny_cloudy_day}).
Then the base solar PV capacity that can be hosted in the system with respect to the static limit is,
\begin{equation}
    P_{\text{pv},i}^{\text{base}}(t) := L_{\text{pv},i}\frac{P_{\text{pv},i}(t) }{\max_{t} \left\{P_{\text{pv},i}(t)\right\}}, 
    \label{eq:P_base}
\end{equation}
 where $P_{\text{pv},i}(t)$ is the PV power at node $i$ at time $t$. Thus, the base PV energy in node $i$ will be,
\begin{equation}
    E_{\text{pv},i}^{\text{base}}= \Delta t\sum_{t=0}^T P_{\text{pv},i}^{\text{base}}(t),
    \label{eq:E_base}
\end{equation}
where $E_{\text{pv}}^{\text{base}}$ represents the total energy generated by the base PV system over the time horizon $T$ (one year) and $\Delta t$ is the time step (5 minutes) used in the calculation. Since the PV capacity is within DHC limits at all times $t$, there is zero curtailment of PV production necessary over the year (i.e., voltage and transformer limits are respected at all time as guaranteed by \textbf{(P1)}). However, if we exceed $L_{\text{pv},i}$, then curtailment is likely expected over a year and focus of next section.

\subsection{Additional solar PV capacity and PV curtailment}
\begin{figure}[!tbp]
    \centering
    \includegraphics[width=0.45\linewidth]{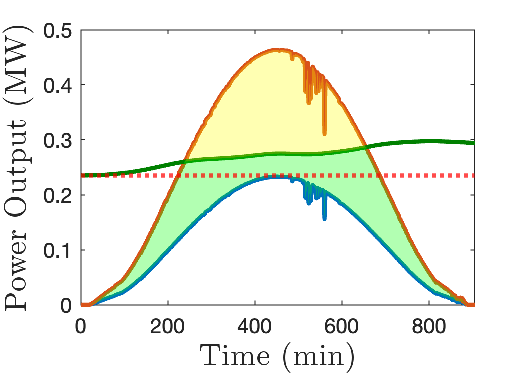}
    \hfill
    \includegraphics[width=0.45\linewidth]{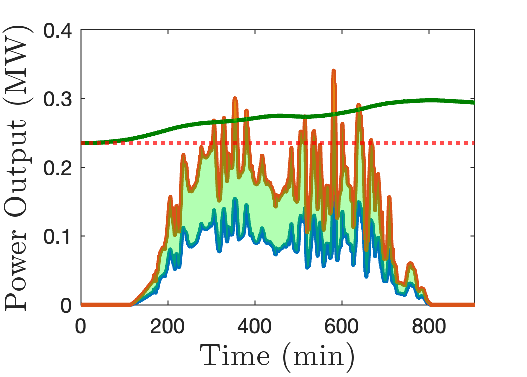}
   \caption{(Left) Representative sunny day for node 35: yellow shaded area represents curtailed energy ($E_{\text{pv},35}^\text{curt}$), green shaded area: additional energy ($E_{\text{pv},35}^\text{add}$), red dotted line: static limits, and the dark green line: dynamic hosting capacity over time computed using Scenario 1-F2. (Right) Representative cloudy day.}
    \label{fig:sunny_cloudy_day}
\end{figure}

Next, we study the effects of deploying additional PV to the grid, focusing on hosting capacity and curtailment, using the six scenarios presented in Section~\ref{sec:Fairness_of_CIA} to compute the DHC. 
As PV capacity increases, curtailment gradually becomes necessary. However, because not all days of the year are sunny, curtailment is less likely on cloudy days, enabling more PV energy to be fed into the grid as shown with green shade in Fig.~\ref{fig:sunny_cloudy_day} (Right).
After the increase, the PV power output (MW) of the system at time $t$ is
\begin{equation}
    P_{\text{pv},i}^{\text{new}}(t) = (1+ \Delta c) P_{\text{pv},i}^{\text{base}}(t),
    \label{eq:P_new}
\end{equation}
where $\Delta c \ge 0$ denotes the \textit{relative capacity increase}. Therefore, the total PV energy at node $i$ will be,
\begin{equation}
    E_{\text{pv},i}^{\text{new}}= 
    \Delta t \sum_{t=1}^T (1+ \Delta c) P_{\text{pv},i}^{\text{base}}(t).
\end{equation}
If solar power exceeds the dynamic limits, curtailment action must be taken to satisfy grid voltage constraints,
\begin{equation}
    E_{\text{pv},i}^{\text{curt}} = \Delta t\sum_{t=0}^T \max \{0, P_{\text{pv},i}^{\text{new}}(t)-p_{g,i}^+(t)\},
    \label{eq:E_curtailed}
\end{equation}
where $E_{\text{pv},i}^{\text{curt}}$ represents the total curtailed PV energy over period $T$. We are particularly interested in quantifying the additional energy generated and injected into the grid \textit{relative to the base case}, where base case corresponds to the minimum base energy across all scenarios. This additional energy is calculated as,
\begin{equation} \label{eq_Eadd}
    E_{\text{pv},i}^{\text{add}} = E_{\text{pv},i}^{\text{new}} - E_{\text{pv},i}^{\text{curt}} - E_{\text{pv},i}^{\text{base}}, 
\end{equation}
which is illustrated by the green shade in Fig.~\ref{fig:sunny_cloudy_day}. 

\begin{figure}[!tbp]
    \centering
    \includegraphics[width=0.45\linewidth]{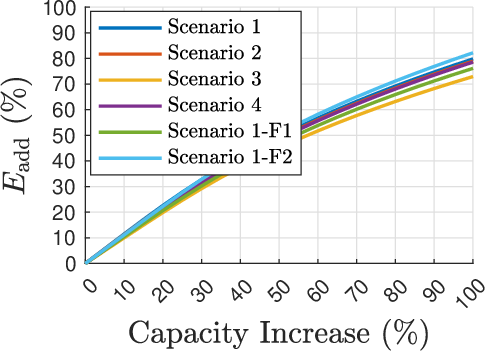}
    \hfill
    \includegraphics[width=0.45\linewidth]{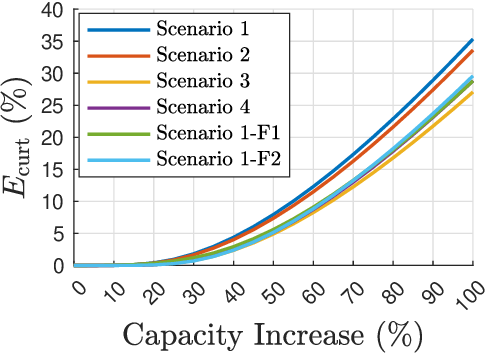}
    \caption{(Left) Total energy added ($E_{\text{pv}}^\text{add}$) versus capacity increase relative to the base case. (Right) Total energy curtailed ($E_{\text{pv}}^\text{curt}$) versus capacity increase relative to the base case.}
    \label{fig:Eadd_and_curt}
\end{figure}

As we increase solar PV capacity, a trade-off emerges between the amount of energy curtailed and the additional energy absorbed by the system. This is illustrated in Fig.~\ref{fig:Eadd_and_curt}, where the additional energy benefit from increased PV capacity decreases monotonically (i.e., the slope in Fig.~\ref{fig:Eadd_and_curt} (Left) is decreasing). Analytically, we can find the limit by rewriting the expression for $E_{\text{pv},i}^{\text{add}}$ in~\eqref{eq_Eadd}, as an explicit function of $\Delta c$ by expanding the different energy terms in~\eqref{eq_Eadd} in terms of $P_{\text{pv},i}^\text{base}(t)$:
\begin{align}
    E_{\text{pv},i}^{\text{add}} := \Delta t \sum^T_{t=0} \min\{\Delta c P_{\text{pv},i}^\text{base}(t), p_{\text{g},i}^+(t) -P_{\text{pv},i}^\text{base}(t) \}
\end{align}
Thus, as $\Delta c \rightarrow \infty$, the additional energy achieves (from below) its limit of
\begin{align}
    E_{\text{pv},i}^{\text{add}} \rightarrow \Delta t \sum^T_{t=0} \left( p_{\text{g},i}^+(t) -P_{\text{pv},i}^\text{base}(t)\right),
\end{align}
which is constant and represents an upper bound on additional energy at each node, and can be described in relative terms as

\begin{align} 
E_{\text{pv},i}^{\text{add,\%}} := \frac{E_{\text{pv},i}^{\text{add}}}{E_{\text{pv},i}^{\text{base}}} \rightarrow 100 \times \left( \frac{\sum_{t=1}^T p_{\text{g},i}^+(t)}{\sum_{t=1}^T P_\text{pv}^\text{base}(t)} - 1 \right).
\end{align}
That is, the maximum additional energy level possible at each node is defined by the ratio of the total DHC at that node (computed from~\textbf{(P1)}) to the total base-case PV power generation at the same node. Of course, with increasing $\Delta c$ and bounded $E_{\text{pv},i}^{\text{add}}$, the PV curtailment will increase, as expected, with increasing PV capacity. 

Interestingly, this analysis highlights why PV curtailment is larger in ``unfair'' scenarios (Scenarios 1 and 2). This is because ``unfair'' allocations assigns disproportionately larger capacities to some nodes, causing $P_{\text{pv},i}^{\text{base}}(t)$ and $P_{\text{pv},i}^{\text{new}}(t)$ to increase significantly at those nodes. 
When we use unfair scenarios, some nodes receive near-zero capacity during peak solar hours, leading to increased curtailment as overall capacity rises as shown in Fig.~\ref{fig:Eadd_and_curt} (Right). Making $ \alpha \propto p_d$ does not improve this situation since nodes with already higher capacities are weighted more heavily in the objective function due to their high demand, further dominating those with lower capacities and intensifying curtailment. 

In fair scenarios, when $\alpha$ is proportional to demand (Scenarios 4 and 1-F2), the total capacity increases more compared to $\alpha = 1_N$ (Scenarios 3 and 1-F1). Since no node is eliminated from participation, ensuring that nodes with higher demands are not allocated less capacity than those with lower demands, thereby promoting equitable allocation. The total additional energy integrated into the grid is higher in unfair scenarios due to their relatively larger capacities compared to fair scenarios as in Fig.~\ref{fig:Eadd_and_curt} (Left). We also observe that curtailment in unfair scenarios grows at a faster rate than the increase in additional energy for these scenarios.

Given a fixed usage charge, $\lambda_\text{curt}$ (\$/kWh),  for any curtailed PV energy, we can estimate the economic cost of curtailment, which varies by utility (e.g., Green Mountain Power (GMP) in Vermont charges about $\$0.20$/kWh~\cite{GMP_rates}). The total cost of curtailment is then,
\begin{equation}
\label{eq:cost_of_curt}
    C_{\text{curt}} = \lambda_{\text{curt}} \sum_{i=1}^NE_{\text{pv},i}^{\text{curt}},
\end{equation}
where $\sum_{i=1}^N E_{\text{pv},i}^{\text{curt}}$ represents the total curtailed energy in the network. Following this, we show how changes in demand significantly impact the HC.
\subsubsection{Demand Sensitivity Analysis}
\begin{figure}[!tbp]
    \centering
    \includegraphics[width=0.45\linewidth]{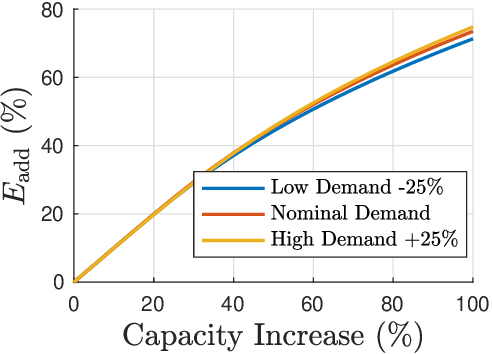}
    \hfill
    \includegraphics[width=0.45\linewidth]{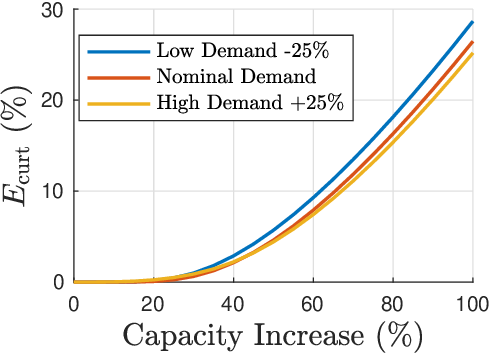}
    \vspace{0.5cm}
    \caption{
        Percentage of (Left) additional energy,
        (Right) curtailed energy
        vs. capacity increase under varying demand scenarios using the allocation Scenario 1-F2.
    }
    \label{fig:Eadd_Ecurt_CT}
\end{figure}

Note that the demand profile affects the HC via \textbf{(P1)}. As shown in Table~\ref{tab:HC_ranges}, HC increases with higher demand and decreases with lower demand, as expected. Therefore, we examine the sensitivity of the results by accounting for the variation in demand using Scenario 1-F2, which results in a fair capacity distribution across nodes as explained in Section~\ref{sec:quantifying_fairness}. Fig.~\ref{fig:Eadd_Ecurt_CT} (Left) shows that with a 25\% increase in demand, the additional energy supplied to the grid also increases due to the expanded range of HC, as indicated in Table~\ref{tab:HC_ranges}. Since the static limits are based on the minimum DHC, a larger range resulting from higher demand leads to greater additional energy. Additionally, curtailment decreases with higher demand.
\begin{table*}[!tbp]
    \centering
    \renewcommand{\arraystretch}{1.2} 
    \caption{Aggregate Hosting Capacity Under Different Demand Profiles and Scenarios}
    \label{tab:HC_ranges}
    \begin{tabular}{rccccccc}
        \toprule
        & \textbf{Scenario 1} & \textbf{Scenario 2} & \textbf{Scenario 3} & \textbf{Scenario 4} & \textbf{Scenario 1-F1} & \textbf{Scenario 1-F2} \\
        \midrule
        \textbf{Low Demand (-25\%)}        & [5.1, 5.7] MW & [4.8, 5.4] MW & [4.2, 5.0] MW & [4.5, 5.3] MW & [4.4, 5.2] MW  & [4.8, 5.6] MW\\ 
        \textbf{Nominal Demand}            & [5.2, 6.1] MW & [5.0, 5.8] MW & [4.4, 5.4] MW & [4.8, 5.8] MW & [4.6, 5.7] MW & [5.0, 6.1] MW\\ 
        \textbf{High Demand (+25\%)}       & [5.4, 6.5] MW & [5.2, 6.2] MW & [4.6, 5.9] MW & [5.0, 6.3] MW & [4.9, 6.1] MW & [5.2, 6.5] MW \\ 
        \bottomrule
    \end{tabular}
\end{table*} 

In the next section, we use the previously discussed additional and curtailed energy from solar PV to estimate environmental carbon and economic revenue impacts. 

\subsection{Carbon Footprint Analysis}
\label{sec:cfp}
The additional clean solar PV energy that is injected into the distribution grid holds benefits associated with reducing CO$_2$ emissions. To quantify these benefits, we consider the economic cost of curtailment (as cost) and the social benefit of reduced carbon (as revenue). While energy costs are well-understood from utility rates (e.g., \$/kWh) in different locations, the (social) cost of carbon varies significantly around the world. In this study, we examine a range of carbon prices from \$50 to \$200 per ton of CO$_2$ (tCO$_2$), recognizing the considerable variability in carbon pricing across policies, regional trading zones, and models\footnote{On the lower end, carbon pricing can be as low as $\text{\$}0.46/\text{tCO$_2$}$~\cite{worldbank_carbonpricing}, while some estimates suggest it could be as high as $\text{\$}185/\text{tCO$_2$}$~\cite{Rennert2022}}.
The life-cycle carbon footprint of solar PV is estimated at $m_{\text{pv}}= 40$~gCO$_2$/kWh, whereas fossil fuel generators like natural gas and coal have footprints ranging from $ m_{\text{fuel}} \in [500, 1000]$ gCO$_2$/kWh depending on altitude, temperature, and fuel types~\cite{nrel_lca}. 
The difference in generation-based emission (i.e., $m_\text{fuel}(t) - m_\text{pv}(t)$) will inform us of the potential carbon benefit from additional solar PV energy. Of course, at different locations and times of day, the grid experiences different (marginal) emissions, $m_\text{grid}(t) \in [40,1000]$  gCO$_2$/kWh, based on which generators are on the margin to meet the next unit of demand. To estimate $m_\text{grid}(t)$, we consider the Marginal Operating Emissions Rate (MOER), which is defined as the amount of carbon emissions associated with a marginal increase or decrease in electricity demand~\cite{watttime_api}. 

\begin{figure}[!tbp]
    \centering
    \includegraphics[width=0.49\linewidth]{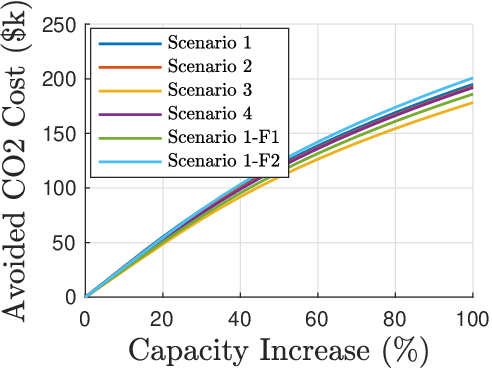}
    \hfill
    \includegraphics[width=0.49\linewidth]{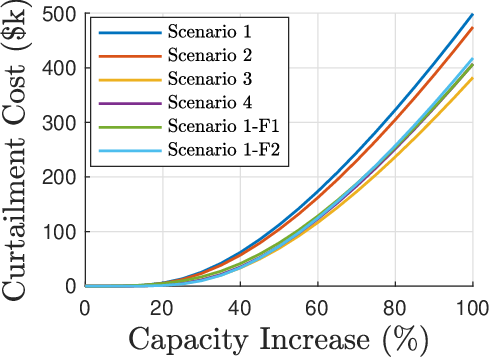}
    \\
    \includegraphics[width=.75\linewidth]{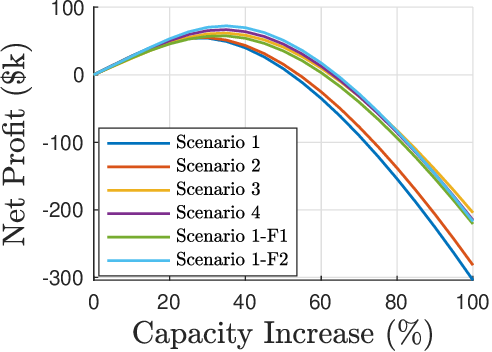}
    \caption{(Top Left) Avoided CO$_2$ costs ($\$k$), (Top Right) Curtailment costs ($\$k$), (Bottom) Net profit ($\$k$) as a function of capacity increase ($\%$) for the six allocation methods with a \$100 carbon price.}
    \label{fig:net_profit}
\end{figure}

Thus, by leveraging the MOER data associated with the US State of Vermont over one year from WattTime's API~\cite{watttime_api}, we can compute the avoided CO$_2$ emissions as,
\begin{equation}
    A_{\text{CO2},i} = \Delta t\sum_{t=1}^{T}P_{\text{pv},i}^{\text{add}}(t)(m_{\text{grid}}(t)-m_{\text{pv}}(t))10^{-6}
\end{equation}
where $P_{\text{pv},i}^{\text{add}}(t) = P_{\text{pv},i}^{\text{new}}(t) - P_{\text{pv},i}^{\text{curt}}(t) - P_{\text{pv},i}^{\text{base}}(t)$ is the additional power at time $t$ and $P_{\text{pv},i}^{\text{curt}}(t) = \max \{0, P_{\text{pv},i}^{\text{new}}(t)-p_{g,i}^+(t)\}$ is the curtailed power at time $t$. $A_{\text{CO$_2$},i}$ represents the tons of CO$_2$ avoided at node $i$ over time period $T$, $m_{\text{grid}}(t)$ is MOER of the grid in gCO$_2$/MWh (reflecting the mix of generation sources), $m_{\text{pv}}(t)$ is the lifetime carbon footprint of solar PV, at time $t$. Therefore, the avoided additional benefit of switching to solar PV is,
\begin{equation}
    C_{\text{rev}} = \sum_{i=1}^NA_{\text{CO$_2$},i} \cdot \lambda_{\text{CO$_2$}},
\end{equation}
where $C_{\text{rev}}$ is the total carbon revenue of the grid that is gained from CO$_2$ offsets enabled by the increased PV capacity and $\lambda_{\text{CO$_2$}} $ is the carbon price. Using six scenarios outlined in Section~\ref{sec:Fairness_of_CIA}, the avoided CO$_2$ costs are shown in Fig.~\ref{fig:net_profit} (Top Left). 

With both the carbon benefits and curtailment costs from~\eqref{eq:cost_of_curt} quantified, we can calculate the utility's net profit from integrating solar PV into the system as follows:
\begin{equation}
\label{eq:net_profit}
    \text{NP} = C_{\text{rev}} - C_{\text{curt}}.
\end{equation}
The simulation results are shown in Fig.~\ref{fig:net_profit} (Top Left) for the six different scenarios when $\lambda_{\text{CO$_2$}} =\$100$/tCO2 and $\lambda_{\text{CO$_2$}} =\$0.20$/kWh for VT. 
Clearly, for large solar PV capacity increases, we expect the unbounded curtailment costs to dominate the net-profits (that are negative) since the carbon benefits from additional PV energy saturates in the limit. However, for the initial PV capacity additions, we expect little to no curtailment costs, so carbon benefits will dominate and net-profits increase (and are positive). Thus, by the Mean Value Theorem, the net profits must achieve a maximum, which occurs at $\Delta c \approx 30\%$ PV capacity increase.

Interestingly, equitable PV nodal distribution of DHC, which we obtain using Scenario 1-F2, leads to higher system-wide benefits, including increased net profits as illustrated in Fig.~\ref{fig:net_profit} (Bottom). This is due to reduced curtailment and increased additional energy. When nodes with lower demands are also included in allocated DHC, they contribute to reducing curtailment and increasing network utilization of solar PV. Key numerical details of Fig.~\ref{fig:net_profit} are summarized in Table~\ref{tab:metrics}.

\begin{table*}[!tbp]
    \centering
    \renewcommand{\arraystretch}{1.2} 
    \caption{Quantifying annual impacts of [5\%, 50\%] increase in PV Capacity across all scenarios}
    \label{tab:metrics}
    \begin{tabular}{rcccccc}
        \toprule
         \textbf{Impacts} & \textbf{Scenario 1} & \textbf{Scenario 2} & \textbf{Scenario 3} & \textbf{Scenario 4} & \textbf{Scenario 1-F1} & \textbf{Scenario 1-F2} \\
        \midrule
        \textbf{$E_{\text{pv}}^{\text{add}} (\%)$}& [5.8, 53.3] & [5.6, 52.6] & [5.0, 48.5] & [5.4, 52.2] & [5.2, 50.5] & [5.6, 54.6] \\ 
        \textbf{$E_{\text{pv}}^{\text{curt}} (\%)$}            & [0.00, 10.0] & [0.0, 9.3] & [0.0, 6.5]  & [0.0, 6.9]  & [0.0, 7.3]  & [0.0, 6.9]  \\ 
        \textbf{Avoided CO2 Cost (\$K)} & [14.1, 130.2] & [13.7, 128.4] & [12.2, 118.5] & [13.1, 127.5] & [12.8, 123.2] & [13.6, 133.3] \\ 
        \textbf{Curtailment Cost (\$K)} & [0.00, 141.74] & [0.00, 131.9] & [0.00, 91.7] & [0.00, 96.9] & [0.1, 102.5] & [0.00, 97.4] \\ 
        \bottomrule
    \end{tabular}
\end{table*}

Next, we examine the wide range of carbon prices to analyze their impact on net profit.
\subsubsection{Impact of Carbon Price on Net Profit}
\begin{figure}[!tbp]
    \centering
    \includegraphics[width=\linewidth]{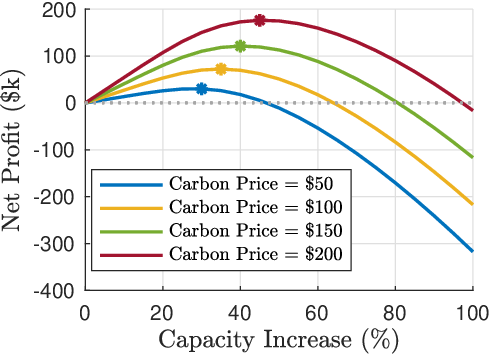}
    \caption{Net profit (\$k) as a function of capacity increase ($\%$) for varying carbon price levels (\$50, \$100, \$150, \$200) using Scenario 1-F2 and nominal demand. Maximum net profits are shown with stars.}
    \label{fig:varying_ct}
\end{figure}

To examine the impact of carbon prices on net profit, we fix the allocation method to Scenario 1-F2, and use the nominal demand profile. As shown in Fig.~\ref{fig:varying_ct}, higher carbon prices significantly improve net profit, reaching the maximum profits in greater capacity increases. This is because higher carbon prices amplify the financial benefits of avoided emissions, offsetting the costs associated with curtailment. Note that this analysis does not account for the costs associated with capacity expansion. Consequently, as the carbon price increases, the net profit starts becoming negative at higher capacity levels as shown in Fig.~\ref{fig:varying_ct}.

Since CO$_2$ emissions vary based on the generation mix of the grid at different locations, in the next section we discuss the locational impact of increasing PV systems on CO$_2$ reduction benefits.

\subsection{Locational Impact on CO2 Reduction Benefits}
\begin{figure}[!tbp]
    \centering
    \includegraphics[width=0.8\linewidth]{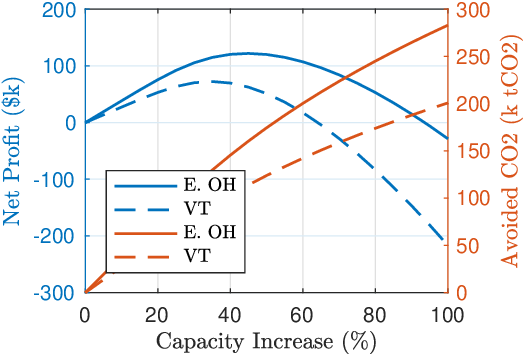}
    \caption{Net profit (\$k, left axis) and avoided CO$_2$ emissions (k tCO$_2$, right axis) as a function of capacity increase ($\%$) for Eastern Ohio and Vermont when price for carbon is $\$100$. }
    \label{fig:VT_vs_OH}
\end{figure}

The carbon benefits will differ across regions due to variations in their energy generation mixes. For instance, a region with a higher share of renewable energy in its grid, such as wind or hydroelectric power, will see fewer benefits from transitioning to solar, as its carbon emissions are already relatively low. Conversely, a state with a generation mix heavily reliant on fossil fuels will experience more significant carbon reductions and financial gains by shifting from fossil-based generation to solar power. Accounting for the cost of curtailing electricity in regions like Eastern Ohio and Vermont, we chose $0.15$/kWh for the Eastern Ohio utility~\cite{EIA_table_5_6_a}, along with their respective MOER data from WattTime~\cite{watttime_api}. Our comparison shows that in states such as Ohio, the implementation of DHC for solar PV delivers transformative carbon reductions in regions like Eastern Ohio, where the reliance on fossil fuels magnifies the environmental benefits of integrating clean energy at scale.
This is illustrated in Fig.~\ref{fig:VT_vs_OH} along with the thousand tons of CO$_2$ avoided by transitioning from fossil-based to solar energy sources.

\section{Conclusion}
\label{sec:conclusion}
This paper improves the CIA approach from \cite{NAWAF2021} by tightening the envelope of branch currents and extending the analysis with a time-series approach to assess system performance over a full day. Leveraging this approach under this study’s fairness definition, we quantified fairness temporally and spatially. Our findings show that maximizing the HC using $\varepsilon$-fair constraint with proportional node weights (Scenario 1-F2) enable equitable grid access. Simulation results show that it is possible, with a relatively small amount of PV curtailment ($\approx 5\%$), to achieve a significant increase ($\approx 50\%$) in the overall solar hosting capacity, offering potential carbon benefits when a carbon price is applied. Additionally, equitable resource distribution leads to higher net profits compared to unfair allocations, benefiting both the economy and the environment. Lastly, the locational benefits of fair PV curtailment highlight the importance of considering the existing energy mix when evaluating renewable energy integration. 

Future work will explicitly consider uncertainty in net-demand at each node of the network for the HC allocation problem. In addition, we are interested in considering geographically diverse solar PV generation profiles at each node. Extending this framework to 3-phase networks will also be valuable. Another direction for future research is to explore how the inclusion of reactive power management can influence the HC of distribution networks, with the potential to increase PV HC by using reactive power regulation to reduce necessary PV curtailment. Furthermore, adopting a multi-period optimization approach could enable the inclusion of temporal fairness in the optimization problem.

\appendices

\section*{Acknowledgment}
This work was supported by the National Science Foundation under Award No. ECCS-2047306.

\ifCLASSOPTIONcaptionsoff
  \newpage
\fi

\bibliographystyle{IEEEtran}
\bibliography{references}

\end{document}